\PassOptionsToPackage{prologue,dvipsnames}{xcolor}
\documentclass[a4paper,11pt]{article}

\pdfoutput=1

\usepackage[utf8]{inputenc}
\usepackage{amsmath}
\usepackage{amssymb}
\usepackage{amsfonts}
\usepackage{mathrsfs}
\usepackage{graphicx}
\usepackage{booktabs,adjustbox}
\usepackage{caption}
\usepackage{slashed}
\usepackage{verbatim}
\usepackage{float}
\usepackage{setspace}
\usepackage{subfig}
\usepackage{jheppub}

\usepackage{color}
\usepackage{ulem}

\usepackage{bookmark}
\usepackage{wasysym}
\usepackage{braket}
\usepackage{dsfont}

\usepackage{enumitem}
\usepackage{physics}
\usepackage{tikz-feynman}

\usepackage{dcolumn}

\allowdisplaybreaks

\def\p{\textbf{p}}
\def\q{\textbf{q}}

\def\0{\textbf{0}}
\def\x{\textbf{x}}

\usepackage{xcolor}                 
\usepackage[dvipsnames]{xcolor} 
\usepackage{pgfplots}               
\pgfplotsset{compat=1.18}           
\usepackage{tikz}                   
\usepackage{caption}                
\usetikzlibrary{plotmarks}
\usetikzlibrary{angles,quotes}
\usetikzlibrary{shapes.geometric, arrows}
\usetikzlibrary{calc}
\usetikzlibrary{fit,positioning}

\definecolor{lightblue_c}{RGB}{102,153,255}
\definecolor{cadmiumorange}{rgb}{0.93, 0.53, 0.18}
\definecolor{applegreen}{rgb}{0.55, 0.71, 0.0}
\definecolor{candypink}{rgb}{0.89, 0.44, 0.48}
\definecolor{hkustyellow}{RGB}{167, 131, 55}
\definecolor{hkustblue}{RGB}{0, 56, 116}
\definecolor{hkustred}{RGB}{209, 51, 59}
\definecolor{APSred}{RGB}{180,55,55}
\definecolor{APSgreen}{RGB}{70,130,100}    
\definecolor{APSblue}{RGB}{65,110,170}     
\definecolor{APSpurple}{RGB}{120,95,140}   
\definecolor{APSorange}{RGB}{220,140,70}   
\definecolor{APSgray}{RGB}{245,245,245}
\definecolor{APSgrav}{RGB}{59,135,123}

\definecolor{APSblueDark}{RGB}{55,95,150}    
\definecolor{APSblueLight}{RGB}{135,200,255} 

\definecolor{APSgreenDark}{RGB}{59,135,123}  
\definecolor{APSgreenLight}{RGB}{160,220,180} 

\definecolor{APSredDark}{RGB}{174,4,39}     
\definecolor{APSredLight}{RGB}{255,150,150}  

\definecolor{APSorangeDark}{RGB}{215,140,75} 
\definecolor{APSorangeLight}{RGB}{255,200,100} 

\definecolor{APSpurpleDark}{RGB}{120,100,145} 
\definecolor{APSpurpleLight}{RGB}{215,180,255} 

\definecolor{APSgrayDark}{RGB}{110,110,110}   
\definecolor{APSgrayLight}{RGB}{245,245,245}  

\makeatletter
\newcommand{\thickhline}{%
	\noalign {\ifnum 0=`}\fi \hrule height 1pt
	\futurelet \reserved@a \@xhline
}
\makeatother

\usepackage{times}
\usepackage{float}

\usepackage{setspace}  

\title{\boldmath  Joint probes of dark matter annihilation from neutrino detectors and CMB targets}

\author[a]{Ruifeng Leng}
\affiliation[a]{School of Physics, Sun Yat-Sen University, Guangzhou 510275, China}
\emailAdd{lengrf@mail.sysu.edu.cn} 
\author[b]{and Shao-Ping Li}
\affiliation[b]{Marietta Blau Institute for Particle Physics, Austrian Academy of Sciences, Dominikanerbastei 16, A-1010 Vienna, Austria}
\emailAdd{Shaoping.Li@oeaw.ac.at}

\abstract{
Dark matter (DM) annihilation into neutrinos provides a promising observational channel targeted by current and forthcoming neutrino detectors. However, the detection of such neutrino fluxes alone cannot uniquely determine their astrophysical or cosmological origin, such as the recent observations from Super-Kamiokande that hint at a small excess of electron antineutrino events. We propose that the effective number of neutrino species and the spectral distortion of the cosmic microwave background (CMB) can serve as complementary observables to probe neutrino signatures from DM annihilation. Using a simple model-independent analysis, we determine the detection windows of these cosmic observables that overlap with the experimental sensitivities from the Super-Kamiokande, Jiangmen Underground Neutrino Observatory, Hyper-Kamiokande, and the Deep Underground Neutrino Experiment, showing that joint probes of large DM annihilation to neutrinos with MeV-GeV masses can be achieved by neutrino detectors and CMB experiments.
}

\begin{document}
\maketitle
\flushbottom
	
\section{Introduction}
\label{sec:intro}

Dark matter (DM) annihilation into neutrinos of the standard model (SM) remains the least constrained channel in current DM indirect detection, which has created a striking platform for several neutrino detectors and telescopes to measure the neutrino fluxes in astrophysical and cosmological scales,  covering a wide range of DM masses in Borexino~\cite{Borexino:2010zht,Borexino:2019wln}, KamLAND~\cite{KamLAND:2011bnd,KamLAND:2021gvi}, Jiangmen Underground Neutrino Observatory (JUNO)~\cite{JUNO:2015zny,Akita:2022lit,JUNO:2023vyz}, Super-Kamiokande (SK)~\cite{Super-Kamiokande:2005wtt,Super-Kamiokande:2008ecj,Super-Kamiokande:2010tar,Super-Kamiokande:2011lwo,Olivares-DelCampo:2017feq}, Hyper-Kamiokande (HK)~\cite{Hyper-Kamiokande:2018ofw}, Deep Underground Neutrino Experiment (DUNE)~\cite{DUNE:2015lol,Capozzi:2018dat}, IceCube~\cite{IceCube:2016zyt,IceCube:2020wum}, and among other things. 

Remarkably, recent data release by combining the SK-gadolinium phase~\cite{Super-Kamiokande:2021the,Super-Kamiokande:2024kcb} hints at a small excess of electron antineutrino events, which can reach a significant level of $2.3\sigma$ when all SK runs are combined~\cite{harada_2024_12726429,santos_2024_13352059,rogly_2024_13351702,beauchene_2024_13351898}. Still, it is to be confirmed experimentally,  and if positive, to be determined whether such an excess comes from diffuse supernova neutrino background~\cite{Kotake:2005zn,Mirizzi:2015eza,Horiuchi:2018ofe} or new physics, especially related to the origin of neutrino masses and DM physics. When it comes to the cosmic origin, such as MeV-scale DM annihilation~\cite{Palomares-Ruiz:2007trf,Ho:2012ug,Boehm:2013jpa,Arguelles:2019ouk,Sabti:2019mhn,Kanemura:2025byi}, it raises an important question of how we can identify such cosmic origins by using observational strategies beyond the terrestrial neutrino detectors. 
 
For DM annihilation with mass from MeV to GeV scale, nevertheless, as pointed out in Ref.~\cite{Kanemura:2025byi}, most of the detection sensitivities in SK, JUNO, HK, and DUNE require a DM annihilation cross section larger than the standard thermal freeze-out value $2.2\times 10^{-26}\lesssim\langle \sigma v\rangle/~[\mathrm{cm}^{3}/\mathrm{s}] \lesssim 8 \times 10^{-26}$~\cite{Steigman:2012nb,Chu:2023jyb}. 
This implies that, for a constant DM annihilation, DM would once be in thermal equilibrium with neutrinos at the neutrino decoupling epoch, but the relic abundance from thermal freeze-out would be insufficient to account for the observed relic density today. This will introduce a DM \textit{density-deficit problem}:  for a large DM annihilation rate into neutrinos to induce observable neutrino fluxes, an intermediate stage of DM production must be present to replenish the DM abundance after thermal freeze-out.
In particular, one would also be confronted by the density-deficit problem if one aims to test the SK antineutrino excess from MeV-scale DM annihilation via the upcoming JUNO and HK experiments~\cite{Granelli:2026bem,Endo:2026upb}.

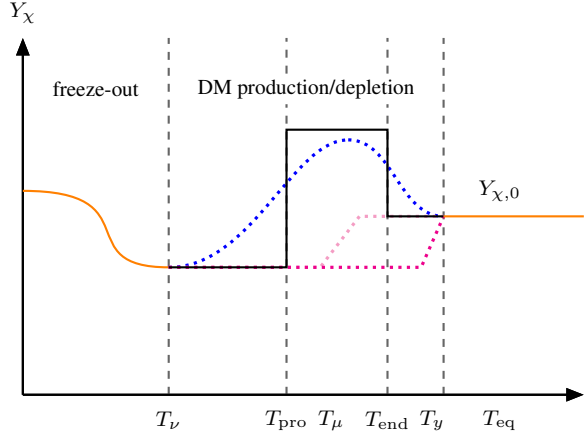
\begin{figure}[t]
\centering
\begin{tikzpicture}[scale = 1]
\begin{axis}[
axis lines = middle,
axis line style={thick, arrows = {-Stealth[inset=0pt]}},  
xtick = \empty,  
ytick = \empty,  
xmin = 0, xmax = 10,
ymin = 0, ymax = 7,
clip = false,
grid = none, 
width = 9cm,
height = 6.3cm,
samples = 300,
legend style = {
at = {(1,1)}, 
font = \scriptsize, 
draw = none, 
fill = none, 
row sep = 3pt,      
cells = {anchor=west}, 
xshift = -10pt,  
yshift = -10pt,    
legend columns = 1,
}
]

\addplot[black!60!white, dashed, thick,forget plot] coordinates {(2.6,0) (2.6,7)}; 
\node at (axis cs:2.6,-0.5) {\scriptsize $T_{\nu}$};

\addplot[black!60!white, dashed, thick,forget plot] coordinates {(7.5,0) (7.5,7)}; 
\node at (axis cs:5.5,-0.5) {\scriptsize $T_{\mu}$};
\node at (axis cs:7.3,-0.5) {\scriptsize $T_{y}$};
\node at (axis cs:8.5,-0.5) {\scriptsize $T_{\rm eq}$};

\addplot[black!60!white, dashed, thick,forget plot] coordinates {(4.7,0) (4.7,5.7)}; 
\addplot[black!60!white, dashed, thick,forget plot] coordinates {(4.7,6.3) (4.7,7)};
\node at (axis cs:4.7,-0.5) {\scriptsize $T_{\rm pro}$};

\addplot[black!60!white, dashed, thick,forget plot] coordinates {(6.5,0) (6.5,5.7)}; 
\addplot[black!60!white, dashed, thick,forget plot] coordinates {(6.5,6.3) (6.5,7)};
\node at (axis cs:6.5,-0.5) {\scriptsize $T_{\rm end}$};

\end{axis}

\begin{axis}[
axis lines = middle,
axis line style={thick, arrows = {-Stealth[inset=0pt]}},  
xtick = \empty,  
ytick = \empty,  
xmin = 0, xmax = 10,
ymin = 0, ymax = 7,
clip = false,
grid = none, 
width = 9cm,
height = 6.3cm,
samples = 300,
legend style = {
at = {(1,1)}, 
font = \scriptsize, 
draw = none, 
fill = none, 
row sep = 3pt,      
cells = {anchor=west}, 
xshift = -10pt,  
yshift = -10pt,    
legend columns = 1,
}
]

\coordinate (start) at (0,4);
\coordinate (dec) at (2.6,2.5);
\draw [orange, thick] (start) .. controls +(2.3,0) and +(-1.8,0) .. (dec);
\node at (axis cs:0,7.5) {\scriptsize $Y_{\chi}$};
\node at (axis cs:1.3,6) {\scriptsize \text{freeze-out}};

\node at (axis cs:5.05,6) {\scriptsize \text{DM production/depletion}};

\coordinate (t1) at (2.6,2.5);
\coordinate (t2) at (7.5,3.5);

\coordinate (ta1) at (5.8,5);
\draw [blue, dotted, very thick] (t1) .. controls +(1.5,0) and +(-1,0) .. (ta1);
\draw [blue, dotted, very thick] (ta1) .. controls +(0.8,0) and +(-0.8,0) .. (t2);
\coordinate (tc1) at (5.3,2.5);
\coordinate (tc2) at (6,3.5);
\draw [Lavender, dotted, very thick] (t1) -- (tc1) -- (tc2) -- (t2);
\coordinate (td1) at (7.1,2.5);
\draw [magenta, dotted, very thick] (t1) -- (td1) -- (t2);
\coordinate (tb1) at (4.7,2.5);
\coordinate (tb2) at (4.7,5.2);
\coordinate (tb3) at (6.5,5.2);
\coordinate (tb4) at (6.5,3.5);
\draw [black, thick] (t1) -- (tb1) -- (tb2) -- (tb3) -- (tb4) -- (t2);

\node at (axis cs:8.5,4) {\scriptsize $Y_{\chi,0}$};
\draw[orange, thick] (7.5,3.5) -- (10,3.5);

\end{axis}
\end{tikzpicture}
\vspace{0.3cm}
\caption{The density-deficit problem in DM annihilation to neutrinos, where the DM evolution with an intermediate stage of production must be present to increase the relic density from the thermal freeze-out value to the present-day value at $T_{\rm eq}\sim 1~\mathrm{eV}$.
Freeze-out may occur as late as the neutrino decoupling temperature, $T_{\nu} \sim 10~\mathrm{keV}$--$1~\mathrm{MeV}$, and the intermediate state takes place at any time between $T_{\nu}$ and $T_{\rm eq}$. The  DM density yield $Y_{\chi}\equiv n_{\chi}/s$ eventually reaches its present-day value, with multiple trajectories illustrating different intermediate production histories for resolving the density-deficit problem.}
\label{fig:DMevo}
\end{figure}

A generic illustration of the density-deficit problem is shown in  Fig.~\ref{fig:DMevo}. DM freeze-out can occur as late as the neutrino decoupling temperature\footnote{For temperatures below the neutrino decoupling $T_{\nu,\mathrm{dec}}\sim2~\mathrm{MeV}$ down to around 50~keV, the ratio of the relic neutrino to photon temperatures continues to evolve because of the photon heating from $e^{+}e^{-}$ annihilation. In this paper, we will simply refer to $T_{\nu}\sim10~\mathrm{keV}$ 
as the temperature at which $e^+e^-$
annihilation is complete, after which the temperature ratio  becomes constant to an excellent approximation.}, $T_{\nu} \sim 10~\mathrm{keV}$--$1~\mathrm{MeV}$~\cite{Mangano:2001iu,Mangano:2005cc,deSalas:2016ztq,Escudero:2018mvt,Bennett:2020zkv,EscuderoAbenza:2020cmq,Cielo:2023bqp}. After freeze-out, an intermediate stage of DM production, possibly followed by depletion, must take place to compensate for the over-annihilated DM density. This may occur during the epochs when primordial spectral distortions of the cosmic microwave background (CMB) could form, such as the $\mu$ distortion at $T_{\mu} \approx 0.47~\mathrm{keV}$ or the $y$ distortion at $T_{y} \approx 12~\mathrm{eV}$~\cite{Burigana:1991eub,Hu:1992dc}. Finally, the density yield $Y_{\chi}$, defined as the DM number density over the entropy density, should reach its present-day value $Y_{\chi,0}$ at the matter-radiation equality epoch $T_{\rm eq}\sim 1~\mathrm{eV}$. In general, observations of neutrino fluxes in neutrino detectors alone cannot uniquely determine the intermediate DM production.

In this work, we propose the use of the effective number of neutrino species $N_{\rm eff}$ and the CMB $\mu$ distortion as potentially complementary observables to probe the patterns of large DM annihilation to neutrinos. We adopt a simple model-independent analysis to determine the detection windows of $N_{\rm eff}$ and the CMB $\mu$ distortion, providing the parameter space for these cosmic observables that can be combined with the observations from neutrino detectors to test the neutrino signals from DM  annihilation. The feasibility of such a combination is based on the following considerations.
After neutrinos fully decouple at $T_{\nu}$, DM annihilation to neutrinos can still generate a significant shift of $N_{\rm eff}$~\cite{Kanemura:2025byi}, leaving impacts on recombination processes that can be detected by using the current CMB experimental techniques.  
On the other hand, when the cosmic temperature cools down to $T_{\mu}$, secondary electromagnetic energy injection from neutrino pair annihilation or coannihilation with the relic neutrino background can perturb the photon bath~\cite{Li:2025clq}. At this stage, photon-number-changing processes, i.e., double Compton scattering and bremsstrahlung, become insufficient under cosmic expansion. 
This can lead to the formation of the primordial CMB $\mu$ distortion~\cite{Burigana:1991eub,Hu:1992dc}, intermediate spectral distortions~\cite{Khatri:2012tw,Acharya:2018iwh} and the $y$ distortion; see also Ref.~\cite{Chluba:2011hw}.  
The current bound from COBE/FIRAS gives: $|\mu|<9\times 10^{-5}, |y|<1.5\times 10^{-5}$~\cite{Mather:1993ij,Fixsen:1996nj}\footnote{Also see Refs.~\cite{Bianchini:2022dqh,Sabyr:2025hwd} that improved the bounds by a factor of 2.}, while the forecast detection limits will be enhanced up to three orders of magnitude from future mission programs, including BISOU~\cite{Maffei:2021xur}, COSMO~\cite{COSMO}, TMS~\cite{TMS}, SPECTER~\cite{Sabyr:2024lgg}, PIXIE~\cite{Kogut:2011xw,Kogut:2024vbi}, super-PIXIE~\cite{Kogut:2019vqh}, and Voyage 2050~\cite{Chluba:2019nxa}. 

 The different trajectories shown in  Fig.~\ref{fig:DMevo} may be induced by some specific particle-physics models such as the neutrinophilic DM scenarios~\cite{Batell:2017cmf,Ballett:2019pyw,Blennow:2019fhy,Li:2022bpp}, but we will adopt in this work a simple model-independent parametrization represented by the black line in Fig.~\ref{fig:DMevo} to simulate the cosmic impacts. While extra contributions to the considered cosmic observables can also appear in the specific model,  such an analysis can only be conducted in a case-by-case way. Instead, our model-independent strategy allows one to focus on the effects entirely from DM annihilation to neutrinos. The resulting conclusions, therefore, should be taken as irreducible contributions for the class of DM models featuring the production history approximated by the black line. In this circumstance, we find that, for light DM at the MeV to $10~\mathrm{GeV}$ scale,  significant observational overlap of $N_{\rm eff}$ and the CMB $\mu$ distortion with the sensitivity of SK, JUNO, HK, and  DUNE exists, illustrating thereby that complementary cosmic signals from DM annihilation into neutrinos beyond the terrestrial neutrino detectors are natural expectations of resolving the DM density-deficit problem.

\section{Neutrino energy release after neutrino decoupling}

\subsection{Simulation of intermediate DM production}
\label{sec:p-wave}
As shown in Fig.~\ref{fig:DMevo}, after DM freezes out from the thermal plasma, its density yield $Y_{\chi}\equiv n_{\chi}/s$ is smaller than the current value $Y_{\chi,0}$ due to a larger annihilation cross section, where $n_{\chi}$ is the DM number density and 
\begin{align}
	s=\frac{2\pi^{2}}{45} g_{s}(T) T^{3} \,,
\end{align}
denotes the SM entropy density with $g_{s}(T)$ the effective number of relativistic degrees of freedom.  
Before the Universe cools down to $T_{\rm eq}$, there must be an intermediate stage to create extra DM abundances. It could occur through hidden particle decay or annihilation from superweakly-interacting massive particles~\cite{Feng:2003xh,Feng:2003uy} or in first-order phase transitions~\cite{Cohen:2008nb,Baker:2019ndr,Wong:2023qon}. 
In general, there are two kinds of evolution for the DM abundance during this intermediate stage. 
It may be gradually increased to accidentally match the present-day value in the end, as shown by the magenta curves in Fig.~\ref{fig:DMevo}. Alternatively, it could first be produced overabundantly, which is then followed by depletion to match $Y_{\chi,0}$, as represented by the blue curve.  

While the precise evolution of $Y_{\chi}$ generally requires a case-by-case treatment, one can adopt suitable approximations to capture the dominant effects. In particular, one may focus on the largest value of $Y_{\chi}$ obtained within a limited time interval, and identify both its typical magnitude and the corresponding period relevant for inducing observable consequences. The estimated $Y_{\chi}$ and the associated time interval can then be mapped onto the parameter space once a specific underlying scenario is considered.  
For a model-independent analysis, we adopt the following parameterization to describe an intermediate DM production:
\begin{align}\label{eq:YDM_param}
Y_{\chi}\equiv 4.37\times 10^{-7}\left(\frac{1~\mathrm{MeV}}{m_{\chi}}\right)\left[f_{\chi} \theta(T_{\rm pro}-T)\theta(T-T_{\rm end})+1\right] \, ,
\end{align}
where $f_{\chi}$ denotes the largest or peak abundance during a short time period $T\in [T_{\rm end},T_{\rm pro}]$. 
The product of the two Heaviside step functions restricts the enhancement to this intermediate stage.
This may correspond to the plateau of the black curve shown in Fig.~\ref{fig:DMevo}. 
Here, $T_{\rm pro}$ denotes the temperature for the DM  production reaching the maximum, and $T_{\rm end}$ denotes the end of  $Y_{\chi}$ residing at the maximum.  

The approximation used in Eq.~\eqref{eq:YDM_param} guarantees that when the production and depletion end at $T<T_{\rm end}$, the relic yield $Y_{\chi}$ would match the present-day value: $\Omega_{\chi}h^{2}\approx 0.12$~\cite{Planck:2018vyg}. Whenever necessary, we will comment on why the parameterization of Eq.~\eqref{eq:YDM_param} is reasonable.  By adopting Eq.~\eqref{eq:YDM_param}, we neglect the subdominant contributions during the production and depletion processes that feature a smaller $f_{\chi}$. 
We will specify the regimes in which neutrino injection in the early Universe will create observable impacts through $N_{\rm eff}$ and the CMB spectral distortion, and identify the regions of parameter space that overlap with the sensitivities of neutrino detectors. 
For this purpose, we should emphasize that Eq.~\eqref{eq:YDM_param} should be treated as a benchmark parameterization that can shed light on the underlying physics for the intermediate  DM production. In addition, all the cosmic impacts considered below entirely come from the effects caused by DM annihilation to neutrinos. Since extra contributions to the cosmic observables may also arise in a specific particle-physics model beyond the pure DM annihilation effects, our results presented in this work should be taken as an irreducible and conservative estimate for patterns of large DM annihilation to neutrinos.

\subsection{Spectrum of neutrino injection}
\label{sec:neu_inj}

Different from long-lived particle decays in the early Universe~\cite{Hambye:2021moy,Li:2023puz,Bianco:2025boy,Li:2025clq}, where dominant neutrino injection only appears at the lifetime scale of the decaying particle, DM annihilation persists even after DM has decoupled from the thermal plasma in the early Universe.
Therefore, any signals from DM annihilation to neutrinos in present-day DM halos will reflect the same process in earlier epochs. This is the case if the DM annihilation rate is independent of the DM velocity. For annihilation scaling with the DM velocity, the resulting impacts are more significant at earlier times, given that the DM velocity not long after freeze-out is much larger than that at present-day typical galaxies. For these velocity-dependent annihilations to neutrinos, however, it implies that the thermally averaged annihilation rate at freeze-out will be much larger than the present-day rate required for observations in neutrino detectors, and hence the DM density-deficit problem becomes even worse. Given this, we will focus on the case of a constant rate for DM annihilation into neutrinos, i.e., the so-called $s$-wave annihilation.

For DM masses in the range of a few MeV to tens GeV, upcoming experiments such as JUNO, HK, and DUNE are expected to improve the sensitivity to the annihilation cross section by about one order of magnitude compared with current bounds. For such low-scale DM masses, constraints from Big-Bang nucleosynthesis (BBN) and the CMB anisotropies are generally weaker than those derived from neutrino injection at energies above the electroweak scale~\cite{Chang:2024mvg,Bianco:2025boy,GrimbaumYamamoto:2025nai}. In particular, electroweak gauge boson cascades are absent for light DM below the electroweak scale, making the associated observational effects from neutrino injection more easily tractable. For heavier DM annihilation, on the other hand, the collision rate would be suppressed by a larger neutrino energy and by a smaller DM number density, such that a significant annihilation rate would typically require larger couplings between the DM and neutrinos, which may introduce more challenges for the class of neutrinophilic DM scenarios. Note that neutrinos injected after the SM neutrino complete decoupling at $T \lesssim 10~\mathrm{keV}$ are hard to rethermalize with the background relic neutrinos when the injected energy is lower than 10~GeV.  This can be quantified by comparing the neutrino interaction rate $\Gamma_\nu$ with the Hubble expansion rate $H$.
The total neutrino cross section follows the standard Fermi-theory scaling $\sigma_{\nu} \sim G_{\rm F}^{2} s$, where $s$ is the Mandelstam variable. For scattering with a thermal background, one has $s \sim E_{\nu} T$ up to an $\mathcal{O}(1)$ angular factor~\cite{Formaggio:2012cpf}. Here $G_{\rm F}\approx 1.167\times 10^{-5}~\mathrm{GeV}^{-2}$ is the Fermi constant.
The interaction rate can then be estimated as 
\begin{align}
\Gamma_{\nu} \sim n_{\nu} \left\langle \sigma_{\nu} v \right\rangle \sim G_F^{2} E_{\nu} T^{4}\,,
\end{align}
where $n_{\nu}\sim T^{3}$ is the number density of background neutrinos.
With the Hubble parameter in the radiation-dominated epoch,
\begin{align}
H=1.66\sqrt{g_{\rho}(T)}\frac{T^{2}}{M_{\rm P}}\,. 
\end{align}
where $M_{\rm P}=1.22\times 10^{19}~\mathrm{GeV}$ is the Planck mass and $g_{\rho}(T)$ is the effective number of relativistic degrees of freedom in the energy density. We find that the injected neutrino energy $E_{\nu}$ should satisfy 
\begin{align}
\left(\frac{E_{\nu}}{10~\mathrm{GeV}}\right)\gtrsim \left(\frac{10~\mathrm{keV}}{T}\right)^{2} \,,
\end{align}
to have $\Gamma_\nu>H$.
This indicates that, although neutrinos are typically decoupled at $T\sim 10~\mathrm{keV}$, the injected neutrinos being sufficiently energetic can still undergo efficient scattering with the background neutrinos.
In particular, for $E_{\nu}\gg 10~\mathrm{GeV}$ at $T\sim 10~\mathrm{keV}$, the neutrino interaction rate can become comparable to or exceed the Hubble expansion rate. In this regime, scattering off the background neutrinos is no longer negligible, and the evolution of injected neutrinos is governed by the combined effects of DM annihilation, subsequent scattering processes, and cosmic expansion.
In contrast, for the DM mass below $10~\mathrm{GeV}$ that will be considered in this work, the injected neutrinos remain out of equilibrium, and scattering between the injected neutrinos and the background plasma can be neglected.
Under this approximation, the injected neutrino distribution can be  determined by the following Boltzmann equation
\begin{align}\label{eq:nu_Boltzmann}
\frac{d f_{\nu}^{\rm inj}}{d  t} =\frac{\partial f_{\nu}^{\rm inj}}{\partial t}-H|\p|\frac{\partial f_{\nu}^{\rm inj}}{\partial |\p|}=\mathcal{C}_{\nu} \,,
\end{align}
where $|\p|$ is the neutrino momentum, and the collision term $\mathcal{C}_{\nu}$ accounts for neutrino production from DM annihilation $\chi \bar{\chi} \to \nu \bar{\nu}$, 
\begin{align}
\mathcal{C}_{\nu}=\frac{1}{2E_{\nu}}\int \left(\prod_{i=\chi, \bar{\chi}, \bar{\nu}} d \Pi_{i} \right) (2\pi)^{4} \delta^{4}(p_{\chi}+p_{\bar{\chi}}-p_{\nu}-p_{\bar{\nu}}) |\mathcal{M}|^{2} f_{\chi} f_{\bar{\chi}}\,,
\end{align}
with the Lorentz-invariant phase-space measure defined as\footnote{We use $d^{3} p$ to denote integration over three-momentum, which should not be confused with the four-momentum inside the Dirac-$\delta$ function.}
\begin{align}
d\Pi_{i} \equiv \frac{d^{3} p_{i}}{(2\pi)^{3}} \frac{1}{2E_{i}} \,.
\label{eq:dPi}
\end{align}
We assume homogeneous and isotropic distribution functions for both neutrinos and DM, such that
$f_{\nu}^{\rm inj}(t,\x,\p)=f_{\nu}^{\rm inj}(t,|\p|)$ and $f_{\chi}(t,\x,\p)=f_{\chi}(t,|\p|)$, as appropriate at the background level in an expanding universe with isotropic DM annihilation.
We consider symmetric DM, for which the particle and antiparticle distribution functions coincide, $f_{\chi}=f_{\bar{\chi}}$. 
The annihilation process is taken to occur after DM has become nonrelativistic, with the amplitude $\mathcal{M}$ evaluated accordingly. 
The coupling of DM to the neutrino flavors is model dependent. For simplicity, we assume here that DM annihilates into the three SM neutrino flavors with approximately equal annihilation rates. In this setup, the total rate of  DM annihilation into neutrinos is given by $3\mathcal{C}_\nu$, and hence the solution $f_{\nu}^{\rm inj}$ denotes the single-flavor distribution of the injected neutrinos.
Finally, we have taken the SM neutrinos to be Dirac fermions for definiteness, but the following conclusions are qualitatively unchanged for Majorana neutrinos.

For the $s$-wave nonrelativistic  DM annihilation, the squared amplitude $|\mathcal{M}|^{2}$ has a simple relation to the thermally averaged cross section\footnote{See Appendix~\ref{app:analy_sol} for more details.},
\begin{align}
	\langle \sigma v\rangle \approx \frac{|\mathcal{M}|^{2}}{32\pi m_{\chi}^{2}}\,,
\end{align}
and the collision rate reduces to 
\begin{align}\label{eq:Cnu-simp}
\mathcal{C}_{\nu}=\frac{2\pi^{2}\langle \sigma v\rangle }{E_{\nu}^{2}}n_{\chi}^{2} \delta(E_{\nu}-m_{\chi})\,.
\end{align}
We then obtain an analytic solution by integrating along the phase-space characteristic:
\begin{align}
f_{\nu}^{\rm inj}(t, |\p|) = \int_{t_{\rm pro}}^{t} d t' \, \mathcal{C}_{\nu} (t', |\p(t')|)=\frac{2\pi^{2}\langle \sigma v\rangle }{m_{\chi}^{3}  H(t_{*})}n_{\chi}^{2}(t_{*}) \theta(t_{*}-t_{\rm pro})\theta(t -t_{*})
\,,
\label{eq:f_nu_sol_0}
\end{align}
where we impose the vanishing initial condition $f_{\nu}^{\rm inj}(t_{\rm pro}, |\p|)=0$, and use the neutrino dispersion relation
\begin{align}
E_{\nu}(t') = |\p| \frac{a(t)}{a(t')} \, ,
\quad
\dot{E}_{\nu}(t') = -|\p| \frac{a(t) \dot{a}(t')}{a^{2}(t')} = -E_{\nu}(t') H(t') \, ,
\end{align}
which defines the resonant time $t_{*}(t,E_{\nu})$ through the condition $E_{\nu}(t_{*}) = m_{\chi}$.
Introducing the dimensionless variables
\begin{align}
	r\equiv \frac{E_{\nu}}{T}\,, \quad x\equiv \frac{T_{\rm pro}}{T}\,,\quad r_{\chi}\equiv \frac{m_{\chi}}{T_{\rm pro}}\,,
\end{align} 
and substituting the parameterization of the intermediate DM yield from Eq.~\eqref{eq:YDM_param} into Eq.~\eqref{eq:f_nu_sol_0}, we arrive at
\begin{align}
f_{\nu}^{\rm inj}(x, r) = 
\frac{2\pi^{2} M_{\rm P}^{*} \langle \sigma v\rangle T_{\rm pro}^{2}}{m_{\chi} r^{4}} \Theta^{2} \theta(r-r_{\chi})\theta(r_{\chi}x -r) \,,
\label{eq:f_nu_sol_1}
\end{align}
where $M_{\rm P}^{*}\approx 0.45 M_{\rm P}/g_{\rho}(T_{*})$.
In general, the effective relativistic degrees of freedom for entropy and energy densities differ, $g_{s}(T) \neq g_{\rho}(T)$. However, after neutrino decoupling, we can adopt $g_{s}(T) \approx g_{\rho}(T) \approx 3.34$, which is a good approximation for the temperature range of interest~\cite{Borsanyi:2016ksw}. 
The two Heaviside step functions in Eq.~\eqref{eq:f_nu_sol_1} have distinct physical origins.
$\theta(r-r_{\chi})$ arises from the redshift of the neutrino energy from the production epoch $T_{\rm pro}$ to the observation temperature $T$, while $\theta(r_{\chi}x -r)$ encodes the kinematic upper bound associated with nonrelativistic DM annihilation.
Given the parameterization of Eq.~\eqref{eq:YDM_param}, an additional Heaviside step function is effectively encoded in the function $\Theta$, which originates from the time (or equivalently temperature) dependence of the DM number density $n_{\chi}$:
\begin{align}
\Theta\equiv 10^{-4}\left(\frac{10~\mathrm{keV}}{T_{\rm pro}}\right)\left[f_{\chi} \theta(r-r_{\chi})\theta(T_{\rm pro}r_{\chi}-T_{\rm end}r)+1\right] \,.
\end{align}
Note that the overall scaling of $f_{\nu}^{\rm inj}$ is independent of the $T_{\rm pro}$ choice, since the explicit dependence cancels in the product $T_{\rm pro}^{2} \Theta^{2}$.


\subsection{Modification to $N_{\rm eff}$}
\label{sec:N_eff}

The neutrino abundance generated by DM annihilation after neutrino decoupling contributes to the radiation energy density and can therefore modify $N_{\rm eff}$:
\begin{align}\label{eq:DeltaNeff}
N_{\rm eff}^{\rm inj}=\frac{8}{7}\left(\frac{11}{4}\right)^{4/3}\left(\frac{2\rho_{\nu}^{\rm inj}}{\rho_{\gamma}}\right),
\end{align}
where $2\rho_{\nu}^{\rm inj}$ denotes the energy density carried by the injected neutrinos and antineutrinos from DM annihilation, and 
\begin{align}
\rho_{\gamma}=\frac{\pi^2}{15} T^4\,,
\end{align} 
is the background photon energy density. 
Here, $N_{\rm eff}^{\rm inj}$ should be understood as the shift of the standard $N_{\rm eff}$ caused by extra radiation energy density.
$\rho_\nu^{\rm inj}$ can be computed from the single-flavor injected distribution $f_{\nu}^{\rm inj}$, i.e., from  Eq.~\eqref{eq:f_nu_sol_1}, via
\begin{align}
\rho_{\nu}^{\rm inj} 
=\sum_{\alpha=e,\mu, \tau}\int \frac{d^{3} p}{(2\pi)^{3}} E_{\nu} f_{\nu_{\alpha}}^{\rm inj}
\approx 3\int \frac{d^{3} p}{(2\pi)^{3}} E_{\nu} f_{\nu}^{\rm inj}
\, ,
\label{eq:rho-nu_inj}
\end{align}
where the last approximation is derived by considering that the DM annihilation rates into three neutrino flavors are of the same order. Unless there is a strong hierarchy among the DM-neutrino flavor couplings, the above result works to a good approximation, owing to the fast redistribution through neutrino oscillations. From Eq.~\eqref{eq:rho-nu_inj}, we arrive at
\begin{align}
N_{\rm eff}^{\rm inj}(T)
\approx 0.0056\left(\frac{\langle\sigma v\rangle}{10^{-24}~\mathrm{cm}^{3}/\mathrm{s}}\right)\left(\frac{1~\mathrm{MeV}}{m_{\chi}}\right)\left(\mathcal{F}_{\rm int}+\mathcal{F}_{\rm rel}\right)
\, ,
\end{align}
where $\mathcal{F}_{\rm int}$ accounts for the contribution from intermediate DM production
\begin{align}
\mathcal{F}_{\rm int}
\equiv f_{\chi}(f_{\chi}+2) \theta \left(T_{\rm pro} - T \right) 
\biggl[ \theta \left(T - T_{\rm end} \right)  \ln \left( \frac{T_{\rm pro}}{T} \right)
+ \theta \left(T_{\rm end} - T \right) \ln \left( \frac{T_{\rm pro}}{T_{\rm end}} \right)
\biggr] 
\, ,
\label{eq:F-int}
\end{align}
while $\mathcal{F}_{\rm rel}$ represents the contribution from relic DM annihilation in the absence of intermediate production\footnote{The contribution from $\mathcal{F}_{\rm rel}$ is consistent with Ref.~\cite{Kanemura:2025byi} after accounting for a total of three neutrino flavors. The small numerical difference derived here is caused by using a more precise DM relic density adopted in Eq.~\eqref{eq:YDM_param}.}
\begin{align}
\mathcal{F}_{\rm rel}
\equiv \theta \left(T_{\rm pro} - T \right) \ln \left( \frac{T_{\rm pro}}{T} \right).
\end{align}
To compare with cosmological observations, we evaluate the above expression at the epoch of photon decoupling, $T = T_{\rm CMB} \!\approx\! 0.26$~eV, which sets the radiation content probed by CMB anisotropies. This yields
\begin{align}
N_{\rm eff}^{\rm inj}
\approx 0.0056\left(\frac{\langle\sigma v\rangle}{10^{-24}~\mathrm{cm}^{3}/\mathrm{s}}\right)\left(\frac{1~\mathrm{MeV}}{m_{\chi}}\right)
\biggl[ f_{\chi}(f_{\chi}+2)  \ln \left( \frac{T_{\rm pro}}{T_{\rm end}} \right)
+ \ln \left( \frac{T_{\rm pro}}{T_{\rm CMB}} \right)
\biggr].
\end{align}

It is worthwhile to mention that since the neutrino injection considered here occurs at $\mathrm{keV}$ temperatures,
after BBN, it can modify $N_{\rm eff}^{\rm CMB}$ inferred at the CMB epoch.
In contrast, $N_{\rm eff}^{\rm BBN}$ denotes the effective radiation content during BBN inferred from the primordial light-element abundances, and therefore remains unchanged by such post-BBN neutrino injection.
Comparing the increasingly stringent BBN constraint with the independent CMB determination therefore
provides a powerful joint test of radiation injection between BBN and
recombination~\cite{Fields:2019pfx,Yeh:2022heq,Yeh:2026pil}.

\section{Formation of CMB spectral distortions from neutrino annihilation}
\subsection{Photon heating from  electron-positron pairs}
\label{sec:ee-pro}
The injected neutrino energy will subsequently be converted into light SM particles once kinematic thresholds are open. 
Given the fact that the number density of injected neutrinos is generally much smaller than that of background thermal relic neutrinos, it has previously been anticipated that pair annihilation of injected neutrinos has a suppressed rate, compared with the coannihilation between background neutrinos and injected neutrinos~\cite{Kawasaki:1994bs,Acharya:2020gfh,Kanzaki:2007pd}.
However, studies on the disintegration of light elements formed at BBN  have shown that pair annihilation of ultrahigh-energy neutrinos can also make a non-negligible contribution to secondary electromagnetic cascades~\cite{Gratsias:1990tr,deLaix:1993gr,Bianco:2025boy}, owing to the larger center-of-mass energy. 
We also note that, after $e^{+}e^{-}$ annihilation, scattering of injected neutrinos off background electrons is strongly suppressed, since charge neutrality implies $n_{e} \simeq n_{b}$, making the relic electron density much smaller than the photon density by the baryon-to-photon ratio $n_{e}/ n_{\gamma}  \simeq  \eta_{b} \sim 10^{-10}$.

For electron-positron pair production through coannihilation $\nu_{\rm inj}+\bar{\nu}_{\rm bg}\to e^{-}+e^{+}$, the injected neutrino energy must satisfy the approximate kinematic threshold
\begin{align}
E_{\nu}^{\rm inj}\gtrsim \frac{m_{e}^{2}}{T_{\nu}^{\rm bg}}\,,
\label{eq:kinthre}
\end{align}
where $T_{\nu}^{\rm bg} \sim T$ is the temperature of the background relic neutrinos, and we have estimated the typical background neutrino energy as $E^{\rm bg}_{\nu}\sim T$. 
In particular, for CMB spectral distortions generated as early as $T_{\mu}=0.47~\mathrm{keV}$~\cite{Burigana:1991eub,Hu:1992dc}, a DM mass larger than roughly $500~\mathrm{MeV}$ is required for coannihilation to proceed efficiently.
If most neutrino injections occur at $T>T_{\mu}$, the neutrino energy would redshift to a smaller value at $T_{\mu}$, such that the DM mass should be larger to open the coannihilation channel at $T_{\mu}$ for generating potential CMB spectral distortions. 
By contrast, for pair annihilation among injected neutrinos,
$\nu_{\rm inj}+\bar{\nu}_{\rm inj}\to e^{-}+e^{+}$, the threshold for converting the injected neutrino energy into an electron-positron pair is much lower,
\begin{align}
E_{\nu}^{\rm inj} \gtrsim m_{e}\,.
\end{align}
This is because both incoming neutrinos are energetic, unlike in the coannihilation channel where the background relic neutrino is soft.

For MeV-scale DM  annihilating into neutrinos, electromagnetic energy transfer can already proceed through pair annihilation of the injected neutrinos.
By contrast, coannihilation in this regime is subject to a strong Boltzmann suppression at $T_{\mu}$, with a factor $\sim e^{-m_{e}/T_{\mu}}$ from the high-energy tail of the thermal relic-neutrino distribution.
In the following, we consider both pair annihilation and coannihilation channels for electron-positron pair production.
We neglect final-state radiation processes that directly produce photons, such as $\nu + \bar{\nu}\to e^{-}+e^{+}+\gamma$, since they are higher order in the electromagnetic coupling.
Although such processes can receive logarithmic enhancements in the infrared regime, the contribution of soft photons to CMB spectral distortions before recombination is expected to be small~\cite{McDonald:2000bk}. 
The produced electron-positron pairs will rapidly lose energy through repeated inverse Compton scattering with background photons, $e^{\pm} +\gamma_{\rm bg}\to e^{\pm}+\gamma$, thereby heating the background photons. 
If the electron-positron pairs are relativistic, only a small fraction of the injected neutrino energy is stored in the electron and positron rest masses. The dominant energy-loss channel is then inverse Compton scattering with background photons~\cite{Blumenthal:1970gc,Chen:2003gz,Padmanabhan:2005es}, whose rate is much larger than the Hubble expansion rate before recombination.
The heated photons subsequently redistribute their energy through Compton scattering with the residual background electrons, $\gamma+e^-_{\rm bg}\to \gamma +e^-$, producing CMB spectral distortions as early as $T=T_{\mu}$.

In general, the energy-transfer rate into electron-positron pairs can be written as 
\begin{align}
\frac{d \rho_{e\bar{e}}}{d t}
=\sum_{\alpha=e,\mu, \tau} \int \left(\prod_{i=\nu_{\alpha},\bar{\nu}_{\alpha}, e, \bar{e}} d \Pi_{i}\right) (E_e+E_{\bar e})
(2\pi)^{4} \delta^{4}(p)
|\mathcal{M}_{\nu_{\alpha}\bar{\nu}_{\alpha}\to e\bar{e}}|^{2}  f_{\nu_{\alpha}} f_{\bar{\nu}_{\alpha}}\,,
\label{eq:drhoee/dt}
\end{align}
where $d \Pi_{i}$ is defined in Eq.~\eqref{eq:dPi}, and $\delta^4(p)\equiv \delta^4(p_{\nu_{\alpha}}+p_{\bar{\nu}_{\alpha}}- p_{e} - p_{\bar{e}})$.
We take the summation of the energy-transfer rate over three neutrino flavors, and have neglected the Pauli-blocking effect for the final-state electron-positron pairs.
For the coannihilation channel, we will take  $f_{\nu_\alpha} f_{\bar{\nu}_\alpha} = f_{\nu_\alpha}^{\rm inj} f_{\bar{\nu}_\alpha}^{\rm bg}$ for definiteness, and then  
 multiply the resulting rate by a factor of two to include the charge-conjugate process $\bar{\nu}_{\rm inj}+\nu_{\rm bg}\to e^{-}+e^{+}$.  
For pair annihilation, both the neutrino distribution functions in Eq.~\eqref{eq:drhoee/dt} are given by Eq.~\eqref{eq:f_nu_sol_1}.
It is worth mentioning again that for temperatures below $\sim 10~\mathrm{keV}$, neutrino oscillations are much faster than the Hubble expansion~\cite{Hannestad:2001iy,Dolgov:2002ab}, and the oscillation time is also much shorter than the relevant scattering time.  Therefore, neutrinos injected by DM annihilation in a given flavor state will instantaneously redistribute among the three flavors. For DM annihilation to three neutrino flavors considered in this work, it is then a good approximation to use Eq.~\eqref{eq:f_nu_sol_1} for each neutrino flavor that participates in the electron-positron pair production.

Using energy conservation, $E_{e}+E_{\bar{e}}=E_{\nu_{\alpha}}+E_{\bar{\nu}_{\alpha}}$, we first perform the final-state electron-positron phase-space integration in the center-of-mass frame, yielding
\begin{align}
\int d \Pi_{e} d \Pi_{\bar{e}} \, (2\pi)^{4} \delta^{4}(p_{\nu_{\alpha}}+p_{\bar{\nu}_{\alpha}}- p_{e} - p_{\bar{e}}) |\mathcal{M}_{\nu_{\alpha}\bar{\nu}_{\alpha}\to e\bar{e}}|^{2}
=\frac{G_{\rm F}^{2} s^{2}}{3\pi}\mathcal{G}_{\alpha}\theta(s-4m_{e}^{2})\,,
\label{eq:int-ee}
\end{align}
where $G_{\rm F}$ denotes the Fermi constant, $s=(p_{\nu_{\alpha}}+p_{\bar{\nu}_{\alpha}})^{2}$ is the Mandelstam variable, and $\theta(s-4m_{e}^{2})$ is the Heaviside step function enforcing the electron-positron production threshold.
The flavor-dependent function $\mathcal{G}_{\alpha}$ is given by
\begin{align}
\mathcal{G}_{\alpha} = \left(1+\frac{2m_{e}^{2}}{s}\right) g_{W}^{\alpha} -\frac{3m_{e}^{2}}{s}\,,
\end{align}
where
\begin{align}
g_{W}^{e} = \left(1+4\sin\theta_{W}^{2}+8\sin\theta_{W}^{4}\right) \,,
\quad
g_{W}^{\mu} = g_{W}^{\tau} = \left(1-4\sin\theta_{W}^{2}+8\sin\theta_{W}^{4}\right) \,,
\end{align}
with $\theta_{W}$ being the weak mixing angle and $\sin^{2}\theta_{W}\approx 0.22$. 
In the present estimate, we include the flavor dependence through the coupling factor $\mathcal{G}_{\alpha}$. 
The $\nu_{e}+\bar{\nu}_{e}\to e^{-}+e^{+}$ channel gives the leading contribution, since it receives both charged-current and neutral-current contributions, while the $\nu_{\mu}\bar{\nu}_{\mu}$ and $\nu_{\tau}\bar{\nu}_{\tau}$ channels proceed only through neutral-current interactions and are suppressed relative to the $\nu_{e}\bar{\nu}_{e}$ channel by $g_{W}^{\mu,\tau}/g_{W}^{e}  \simeq  0.22$.

Next, it becomes straightforward to complete the calculation of the energy-transfer rate by applying the neutrino distribution functions.  For the coannihilation channel, the relic neutrino background is described by the Fermi-Dirac distribution
\begin{align}
f_{\nu}^{\rm bg}(E_{\nu})=\frac{1}{e^{E_{\nu}/T_{\nu}^{\rm bg}}+1}\,,
\end{align}
and we neglect the small difference between the three flavors. Furthermore, we take, for simplicity, the neutrino temperature to be the one following the standard neutrino decoupling $T_{\nu}^{\rm bg} =(4/11)^{1/3} T$. While the modification of $N_{\rm eff}$ in general induces a shift of this temperature ratio, the impact is small when evaluating the energy-transfer rate. 
For pair annihilation, on the other hand, both of the two neutrino distribution functions are given by Eq.~\eqref{eq:f_nu_sol_1}. Because the integrand contains several kinematic step functions, we will evaluate Eq.~\eqref{eq:drhoee/dt} numerically.

\subsection{The CMB $\mu$ distortion}
\label{sec:p-channel-mu}

In this section, we will consider the formation of the CMB $\mu$ distortion, which can be accurately described by using semi-analytic methods.  
The extension to the formation of the $y$ distortion and intermediate-type distortions~\cite{Khatri:2012tw,Acharya:2018iwh} can also be taken, but requires full numerical solutions of the spectral-distortion equations. Nevertheless, we should mention that, 
since both of the intermediate-type distortions and the $y$ distortion are formed at a later stage, intermediate DM production at these later epochs could lead to stronger impacts at the matter-radiation equality and recombination epochs, suffering more severe constraints from large-scale structure formation and CMB anisotropies. For intermediate DM production at much earlier times, on the other hand, the generation of these intermediate-type and $y$ distortions would require heavier DM since the redshift effect of the injected neutrino energy is stronger.

The heating rate of the background photons induced by the kinetic energy of the electron-positron pairs can then be approximated as
\begin{align}
\frac{d\rho_{\gamma}}{dt}\approx \frac{d\rho_{e\bar{e}}}{dt}\,.
\label{eq:drho_gamma_dt}
\end{align}
Although not all of the energy stored in the electron-positron pairs is converted into photon energy, Eq.~\eqref{eq:drho_gamma_dt} provides a good approximation for predicting the resulting $\mu$ distortion for relativistic $e^\pm$~\cite{Blumenthal:1970gc,Chen:2003gz,Padmanabhan:2005es,Li:2024xlr}.  
After the subsequent Comptonization, the transferred photon energy generates a primordial $\mu$ distortion in the background photon distribution during the epoch  $12~\mathrm{eV}\lesssim T\lesssim0.47~\mathrm{keV}$, corresponding to the redshift  $z_y\equiv 5\times 10^{4}\lesssim z\lesssim 2\times 10^6\equiv z_{\mu}$~\cite{Burigana:1991eub,Hu:1992dc,Chluba:2011hw}, 
which will survive to the present day. 
The well established analytic formula yields~\cite{Sunyaev:1970er,Khatri:2012tv,Li:2024xlr}
\begin{align}
\mu\approx 1.4 \int_{0}^{\infty}\mathcal{J}_\mu(z)\frac{1}{\rho_{\gamma}}\frac{d\rho_{\gamma}}{dt} \left|\frac{dt}{dz}\right|dz\,,
\label{eq:mu-dis}
\end{align}
where $\mu$ denotes the dimensionless chemical-potential parameter characterizing the CMB spectral distortion. 
$\rho_{\gamma}$ in the denominator is the unperturbed blackbody photon energy density, which serves as the normalization for the injected energy.
$d\rho_{\gamma}/dt$ is the rate of the electromagnetic energy-injection into the photon bath. 
We use $|dt/dz|\approx7.3\times 10^{37}/(1+z)^{3}~\mathrm{keV}^{-1}$, and  $\mathcal{J}_\mu(z)$ is the visibility function for the epoch during which the $\mu$ distortion is formed~\cite{Chluba:2013kua,Chluba:2016bvg}
\begin{align}
\mathcal{J}_\mu(z)=e^{-(z/z_{\mu})^{5/2}}\theta(z-z_{y})\,.
\label{eq:mu-vis}
\end{align}
The exponential factor above accounts for the erasure of distortions at sufficiently high redshift, where photon-number-changing processes efficiently restore a blackbody spectrum, while the step function restricts the formation epoch to the $\mu$-distortion era.

\begin{figure}[t]
\centering
\includegraphics[scale=0.5]{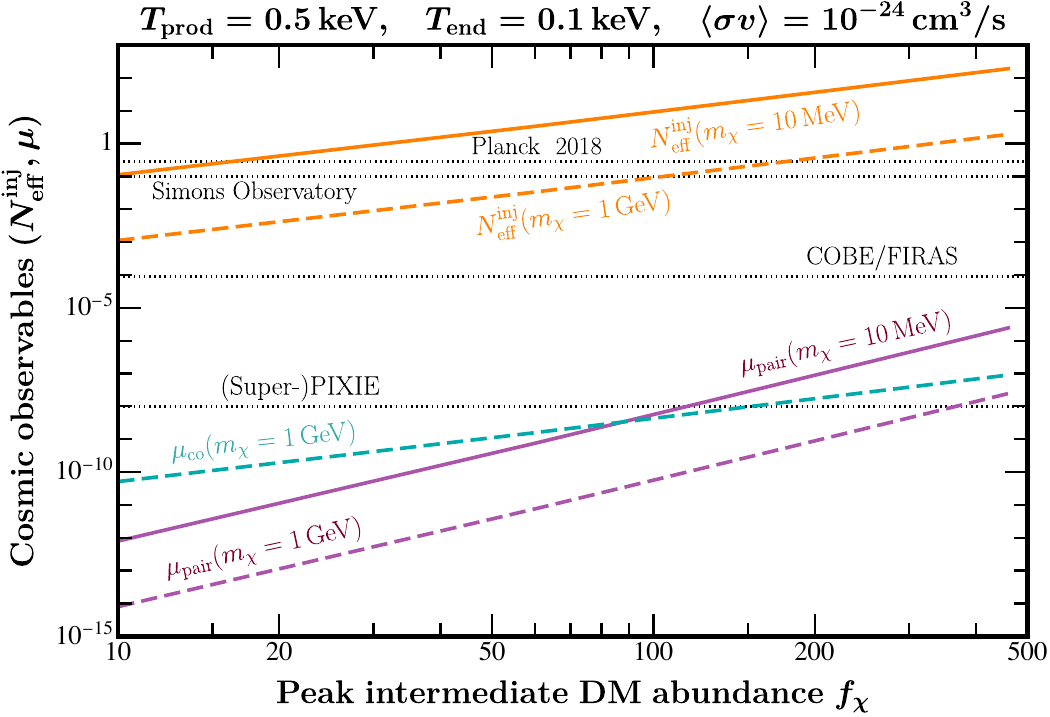} 
\caption{Modification of $N_{\rm eff}$, denoted by $N_{\rm eff}^{\rm inj}$, and the CMB $\mu$ distortion as functions of the peak intermediate DM abundance $f_\chi$. The coannihilation and pair annihilation contributions to the $\mu$ distortion are denoted by $\mu_{\rm co}$ and $\mu_{\rm pair}$, respectively. Injected neutrinos arise from DM annihilation during $T\in[0.1,0.5]~\mathrm{keV}$, with a reference annihilation cross section near the projected sensitivities of JUNO~\cite{JUNO:2015zny,Arguelles:2019ouk,Akita:2022lit}, HK~\cite{Horiuchi:2008jz,Hyper-Kamiokande:2018ofw,Bell:2020rkw}, and DUNE~\cite{Honda:2015fha,Arguelles:2019ouk}. 
For comparison, we show the COBE/FIRAS bound $|\mu| < 9\times10^{-5}$~\cite{Mather:1993ij,Fixsen:1996nj}, a representative future sensitivity $|\mu| \simeq 10^{-8}$ for (Super-)PIXIE~\cite{Kogut:2024vbi}, the Planck limit $N_{\rm eff}^{\rm inj} \lesssim 0.285$~\cite{Planck:2018vyg}, and the projected Simons Observatory sensitivity $N_{\rm eff}^{\rm inj} \simeq 0.1$~\cite{SimonsObservatory:2018koc,SimonsObservatory:2019qwx}.} 
\label{fig:Neffmu-fDM}
\end{figure}

The integration in Eq.~\eqref{eq:mu-dis} can be done numerically by substituting the energy-deposition rate obtained in Eq.~\eqref{eq:drhoee/dt}. 
In Fig.~\ref{fig:Neffmu-fDM}, we show the resulting $N_{\rm eff}^{\rm inj}$ together with the $\mu$ distortion induced by coannihilation, $\mu_{\rm co}$, and by pair annihilation, $\mu_{\rm pair}$.
We fix the annihilation cross section to be $\langle \sigma v\rangle=10^{-24}~\mathrm{cm}^{3}/\mathrm{s}$, which is of the
typical order targeted by upcoming neutrino detectors such as JUNO~\cite{JUNO:2015zny,Arguelles:2019ouk,Akita:2022lit},
HK~\cite{Horiuchi:2008jz,Hyper-Kamiokande:2018ofw,Bell:2020rkw}, and
DUNE~\cite{Honda:2015fha,Arguelles:2019ouk}; see also Ref.~\cite{Arguelles:2019ouk}. 
We take the intermediate DM production to occur over the relatively short temperature interval $T\in[0.1,0.5]~\mathrm{keV}$. Enlarging the interval over which the peak abundance $f_{\chi}$ is maintained does not significantly modify the
predictions for either $N_{\rm eff}^{\rm inj}$ or the $\mu$ distortion.

The reason for the rather weak sensitivity of both $N_{\rm eff}^{\rm inj}$ and the $\mu$ distortion on the intermediate interval temperature can be understood as follows. During the period in which $f_{\chi}$ is active, $N_{\rm eff}^{\rm inj}$ depends only logarithmically on the duration of the
production epoch, $N_{\rm eff}^{\rm inj}\propto \ln(T_{\rm pro}/T_{\rm end})$, as can be inferred from Eq.~\eqref{eq:F-int}.
Therefore, increasing the range of temperatures over which the intermediate abundance persists only
leads to a mild change in $N_{\rm eff}^{\rm inj}$.
By contrast, for pair annihilation, the integrand in Eq.~\eqref{eq:mu-dis} scales approximately as $T$. 
This follows from the scaling of the injected neutrino distribution $f_{\nu}^{\rm inj} \propto T^{4}$, given in Eq.~\eqref{eq:f_nu_sol_1}, together with $1/\rho_{\gamma} \propto T^{-4}$ and $|dt/dz| \propto T^{-3}$. 
For coannihilation with the neutrino background, the relic neutrino distribution is Boltzmann suppressed at low temperatures,
$f_{\nu}^{\rm bg} \propto e^{-m_e/T}$. This exponential suppression strongly reduces the coannihilation contribution to the integrand in Eq.~\eqref{eq:mu-dis} at late times.
As a result, the contribution to the $\mu$ distortion is mainly controlled by energy deposition near the high-temperature end of the $\mu$-distortion era, around $T_{\mu}$. Extending the intermediate DM production period to lower temperatures adds only a subdominant contribution to the integral, and hence has little impact on the predicted $\mu$ distortion.

Conversely, if the intermediate DM production period is made very short, the effects on both observables are suppressed by the small available temperature interval. For $N_{\rm eff}^{\rm inj}$, the logarithmic factor $\ln(T_{\rm pro}/T_{\rm end})$ becomes small when $T_{\rm pro}$ and $T_{\rm end}$ are close. For the $\mu$ distortion, the integrated energy deposition in Eq.~\eqref{eq:mu-dis} is likewise reduced because the integration range over which the source is active is narrow.
Thus, in this limiting case, the impacts of intermediate production on both $N_{\rm eff}^{\rm inj}$ and the $\mu$ distortion are correspondingly suppressed. Nevertheless, for typical intermediate DM production, we expect that the peak abundance would not sharply evolve to vanishingly small values.  These observations in turn justify the proper parameterization of the intermediate DM production via the black line shown in Fig.~\ref{fig:DMevo} or equivalently via Eq.~\eqref{eq:YDM_param}.

Two representative mass choices, $m_{\chi}=10~\mathrm{MeV}$ and $m_{\chi}=1~\mathrm{GeV}$, are shown in Fig.~\ref{fig:Neffmu-fDM}. 
For the lower mass case, the coannihilation contribution to the $\mu$ distortion is strongly suppressed. 
As discussed in Section~\ref{sec:ee-pro}, this suppression originates from the exponential feature of the high-energy thermal tail of the background neutrino distribution required to satisfy the kinematic threshold. 
For $m_{\chi}=1~{\rm GeV}$, this suppression is alleviated, and the coannihilation contribution becomes much larger. 
In this case, $\mu_{\rm co}$ even exceeds $\mu_{\rm pair}$, indicating that coannihilation dominates the generation of the $\mu$ distortion for larger DM masses.
However, this comparison can change if the peak intermediate DM abundance $f_{\chi}$ is increased. 
Since $\mu_{\rm pair}\propto f_{\chi}^{4}$, whereas $\mu_{\rm co}\propto f_{\chi}^{2}$, the pair annihilation contribution grows
more rapidly with $f_{\chi}$ and can dominate for sufficiently large $f_{\chi}$. 
A similar situation will occur for neutrino injection from relic particle decay, where pair annihilation among the neutrinos produced by the decays can become more important than coannihilation if the abundance of the relic particles at decay is sufficiently large~\cite{Li:2025clq}.

Nevertheless, Fig.~\ref{fig:Neffmu-fDM} allows us to draw a general conclusion that $\mu_{\rm pair}$ cannot reach the forecast $\mu$-distortion sensitivity without violating the $N_{\rm eff}$ constraint.
For both benchmark masses, $m_{\chi}=10~\mathrm{MeV}$ and $m_{\chi}=1~\mathrm{GeV}$, we see clearly that $N_{\rm eff}^{\rm inj}$ reaches the current Planck bound when $\mu_{\rm pair}$ is still too small to reach the detection limit. 
In contrast, the coannihilation contribution $\mu_{\rm co}$ can reach the detection limit while $N_{\rm eff}^{\rm inj}$ coincidentally reaches the projected sensitivity of future probes, e.g., Simons Observatory~\cite{SimonsObservatory:2018koc,SimonsObservatory:2019qwx}, provided that the DM mass is above $500~\mathrm{MeV}$.  

\section{Neutrino detectors meet the  targets for the early Universe}
\label{sec:analysis}

\begin{figure}[t]
\centering
\includegraphics[scale=0.41]{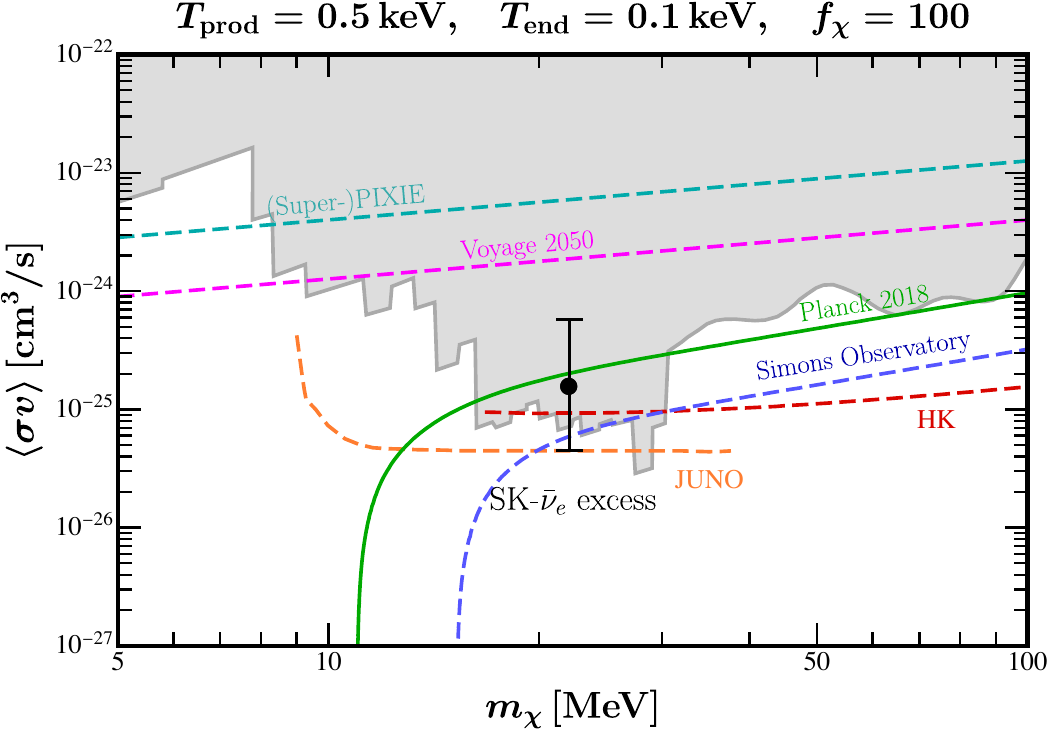}\quad
\includegraphics[scale=0.41]{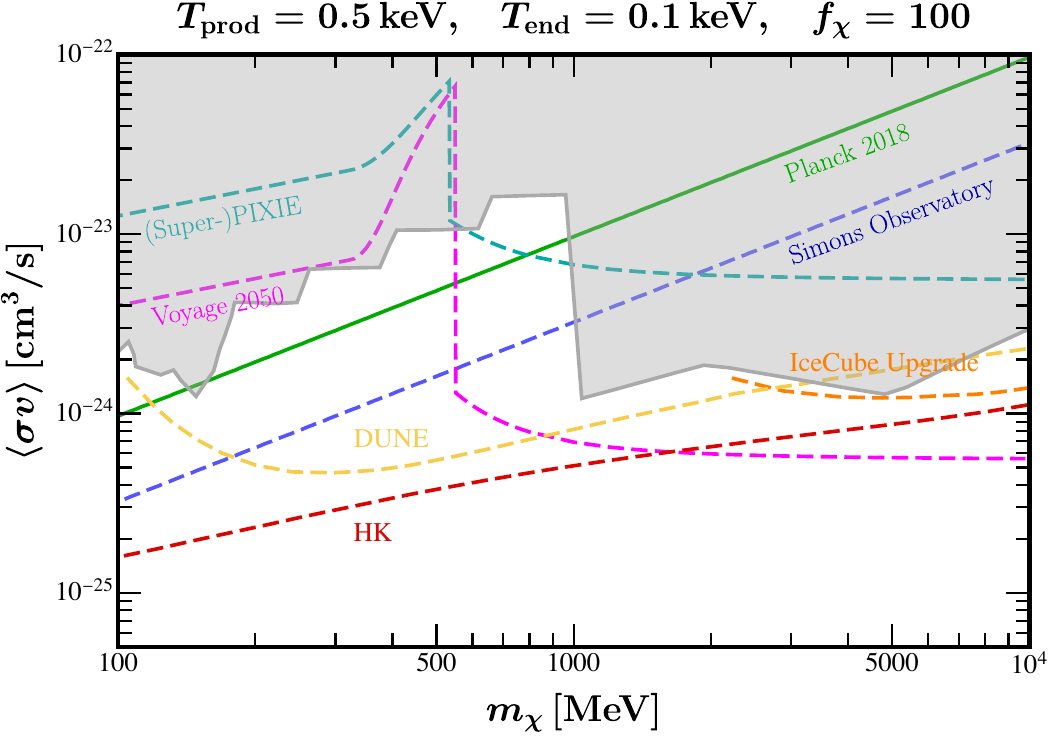} 
\caption{
Detection windows for $N_{\rm eff}^{\rm inj}$, the CMB $\mu$ distortion, and neutrino-detector searches in the $(m_{\chi},\langle\sigma v\rangle)$ plane, where $\langle\sigma v\rangle$ is the total DM annihilation cross section into all three SM neutrino flavors~\cite{Arguelles:2019ouk}. The cosmological curves are labeled by the corresponding limits or sensitivities: Planck 2018 for $N_{\rm eff}^{\rm inj} \simeq 0.285$, Simons Observatory for $N_{\rm eff}^{\rm inj} \simeq 0.1$, (Super-)PIXIE for $|\mu| \simeq 10^{-8}$, and Voyage 2050 for $|\mu| \simeq 10^{-9}$. Dashed curves denote future sensitivities, while solid curves and shaded regions denote current limits. 
Current neutrino-detector exclusions at 90\% C.L. limits are from Borexino~\cite{Borexino:2010zht,Borexino:2019wln}, KamLAND~\cite{KamLAND:2021gvi}, and SK~\cite{Super-Kamiokande:2005wtt,Super-Kamiokande:2008ecj,Super-Kamiokande:2010tar,Super-Kamiokande:2011lwo,Super-Kamiokande:2013ufi}.
In the left panel, the black dot marks the SK-$\bar{\nu}_{e}$ best-fit point for the antineutrino excess, and 
the black error bar shows the corresponding $2 \sigma$ region~\cite{Granelli:2026bem}. 
}
\label{fig:sigmav-mX}
\end{figure}

The modification of $N_{\rm eff}$ and the generation of the CMB $\mu$ distortion with $\langle \sigma v\rangle=10^{-24}~\mathrm{cm}^{3}/\mathrm{s}$ suggest that joint CMB measurements and neutrino detectors can probe overlapping regions of parameter space with large rates of DM annihilation into neutrinos.
To illustrate this point, we show in Fig.~\ref{fig:sigmav-mX} the detection windows in the $(m_{\chi},\langle\sigma v\rangle)$ plane, adopting the benchmark values $f_{\chi}=100$, $T_{\rm pro}=0.5~\mathrm{keV}$, and $T_{\rm end}=0.1~\mathrm{keV}$. The left panel corresponds to the low-mass regime ($m_{\chi}<100~\mathrm{MeV}$), while the right panel shows the high-mass regime ($m_{\chi}>100~\mathrm{MeV}$).
The experiment-labeled curves correspond to target values of the cosmic observables: $N_{\rm eff}^{\rm inj}\simeq0.285, 0.1$ for Planck 2018 and Simons Observatory, respectively, and $|\mu|\simeq10^{-8}, 10^{-9}$ for (Super-)PIXIE and Voyage 2050.

In the left panel of Fig.~\ref{fig:sigmav-mX}, the region between the Planck 2018 and Simons Observatory curves defines the target window $N_{\rm eff}^{\rm inj}\in[0.1,0.285]$, which overlaps appreciably with the parameter space accessible to JUNO and HK for $m_{\chi} \in [10,100]~\mathrm{MeV}$.
By contrast, the projected CMB $\mu$-distortion sensitivities, $|\mu| \in [10^{-8}, 10^{-9}]$, require much larger annihilation cross sections in this mass range.
They are therefore well separated from both the $N_{\rm eff}^{\rm inj}$ target window and the JUNO/HK sensitivities, indicating that the CMB $\mu$ distortion is not expected to be observable in the MeV-scale region.
We  show the current SK-$\bar{\nu}_{e}$ exclusion region for $9~\mathrm{MeV}\lesssim m_\chi\lesssim 30$~MeV, which is evaluated at 90\% C.L. for the normalized Galactic $J$-factor $\mathcal{J}_{\rm avg}=4$~\cite{Arguelles:2019ouk,Granelli:2026bem}.
The updated SK-$\bar{\nu}_{e}$ data by combining all the SK runs, especially including the SK-gadolinium phase, prefer a best-fit point for the antineutrino excess at $m_{\chi} = 22.1~\mathrm{MeV}$, with the black error bar indicating the corresponding $2 \sigma$ range $\mathcal{J}_{\rm avg} \langle\sigma v\rangle \in [1.8 \times 10^{-25}, 2.3 \times 10^{-24}] \mathrm{cm}^{3}/\mathrm{s}$~\cite{Granelli:2026bem}.
We should mention here that a small portion of the $2 \sigma$ SK-$\bar{\nu}_{e}$ excess region still remains below the current exclusion bound.

The SK-$\bar{\nu}_{e}$ constraint requires more accurate treatment for the astrophysical uncertainties, since it is important to confirm and test the potential SK-$\bar{\nu}_{e}$ excess in the MeV-mass regime, and to determine the extent to which the $N_{\rm eff}^{\rm inj}$ target window will overlap with the future reach of JUNO and HK. 
Note that the explanation of the SK-$\bar{\nu}_{e}$ excess does not necessarily lead to the DM density-deficit problem if one adopts the $2\sigma$ level of the observed events, as can be seen from the left panel of Fig.~\ref{fig:sigmav-mX}. The corresponding annihilation cross section can reach down to $\langle\sigma v\rangle \approx 5 \times 10^{-26}\mathrm{cm}^{3}/\mathrm{s}$, comparable with the thermally averaged one predicted by the standard freeze-out. Nevertheless, this value matches the projected sensitivity of JUNO marginally and lies below that of HK. Therefore, the full confirmation and test of the SK-$\bar{\nu}_{e}$ excess through both JUNO and HK would introduce the density-deficit problem. 

We arrive at the conclusion that a possible SK-$\bar{\nu}_{e}$ excess, future neutrino-detector signals, and a measurable shift in $N_{\rm eff}$ can probe a common region of MeV-scale DM annihilation into neutrinos. 
Note that the DM masses below $10~\mathrm{MeV}$ are also constrained by BBN and CMB observations through their impact on neutrino decoupling~\cite{Palomares-Ruiz:2007trf,Ho:2012ug,Boehm:2013jpa,Nollett:2014lwa,Escudero:2018mvt,Sabti:2019mhn,EscuderoAbenza:2020cmq,Sabti:2021reh,Chu:2022xuh}. However, the latest data releases from DESI~\cite{DESI:2024mwx,DESI:2025ejh}, SPT-3G~\cite{SPT-3G:2024atg}, and ACT~\cite{ACT:2025tim} leave the lower mass bound inconclusive~\cite{Kanemura:2025byi}. 
Moreover, depending on the origin of the intermediate DM production, a significant contribution to $N_{\rm eff}$ can arise even for $m_{\chi}>10~\mathrm{MeV}$, where it may dominate over the modification of neutrino decoupling.
Therefore, a signal of $N_{\rm eff}^{\rm inj}$ can be correlated with extra neutrino fluxes observable at SK, JUNO, and HK for $m_{\chi}\in[10,100]~\mathrm{MeV}$.

In the right panel of Fig.~\ref{fig:sigmav-mX}, we show the detection windows for the higher DM mass range $m_{\chi}\in[100,10^{4}]~\mathrm{MeV}$. 
The target region between the Planck 2018 and Simons Observatory curves overlaps with the parameter space accessible to HK and DUNE.
Moreover, for $m_{\chi} \gtrsim 500~\mathrm{MeV}$, the region between the (super-)PIXIE and Voyage 2050 curves also enters the HK/DUNE reach, indicating that the same parameter space can be jointly probed by an observable $\mu$ distortion and $N_{\rm eff}^{\rm inj}$.
For $m_{\chi} \gtrsim 1~\mathrm{GeV}$, we notice that the predicted $\mu$ distortion depends only weakly on the DM mass, while the corresponding $N_{\rm eff}^{\rm inj}$ lies below $0.1$.
Indeed, for $m_{\chi} \gtrsim 500~\mathrm{MeV}$, we can verify that the coannihilation rate in Eq.~\eqref{eq:drhoee/dt} becomes nearly independent of the DM mass. 
It implies that, for heavier DM, the CMB $\mu$ distortion can become the dominant cosmological probe of large annihilation rates, where substantial overlap with the regions accessible to HK and DUNE exists. This behavior is qualitatively consistent with the fact that CMB spectral distortions can be more sensitive than $N_{\rm eff}$ to ultrahigh-energy neutrino injection from superheavy DM or long-lived particle decay~\cite{Acharya:2020gfh}.

We should mention that $|\mu| \in [10^{-8}, 10^{-9}]$ opens a larger window for $m_{\chi} \gtrsim 500~\mathrm{MeV}$ than for $m_{\chi}\in[100,500]~\mathrm{MeV}$. This follows from the different scaling behaviors of the dominant energy-injection channels.
For $m_{\chi}\in[100,500]~\mathrm{MeV}$, the CMB $\mu$ distortion is mainly induced by pair annihilation of injected neutrinos, whose rate scales approximately as $\langle\sigma v\rangle^{2}$. 
Thus, improving the $\mu$ sensitivity by one order of magnitude lowers the required cross section by roughly $\sqrt{0.1}$. 
For $m_{\chi} \gtrsim 500~\mathrm{MeV}$, coannihilation with the relic neutrino background becomes dominant, the rate of which scales approximately linearly with $\langle\sigma v\rangle$. The same sensitivity improvement of the $\mu$ distortion then lowers the required cross section by about $0.1$, producing the larger window between the (Super-)PIXIE and Voyage 2050 curves in the high-mass region.

It should also be mentioned that the kinematic threshold $m_{\chi}\gtrsim 500~\mathrm{MeV}$ follows from Eq.~\eqref{eq:kinthre} for the benchmark choice $T_{\rm pro}=0.5~\mathrm{keV}$. If both $T_{\rm pro}$ and $T_{\rm end}$ are higher than $T_{\mu}$, the redshift of the injected neutrino energy becomes important, and the kinematic threshold for coannihilation shifts to larger DM masses.
In the limit $T_{\rm pro}, T_{\rm end} \gg T_{\nu} \sim 1~\mathrm{MeV}$, injected neutrinos rapidly thermalize with the background plasma, so they can no longer be treated as a late-time contribution to $N_{\rm eff}$ and do not form CMB spectral distortions.

As a final remark, we emphasize that the constant DM annihilation considered in this work corresponds to a DM-velocity independent annihilation cross section. The overlap between the cosmological detection windows and the neutrino-detector sensitivities found in this work depends sensitively on the intermediate DM abundance, parameterized by $f_{\chi}$.
Since $f_{\chi}$ encodes the production history, it can be used to probe the underlying particle-physics model.
A joint measurement or exclusion by CMB observables and neutrino detectors would probe not only the large rates of DM annihilation into neutrinos, but also the mechanism responsible for generating and maintaining the intermediate DM abundance.
We also note that the DM density-deficit problem may be relaxed in velocity-dependent annihilation.  Generally, we may parameterize the thermally averaged annihilation cross section in the partial-wave method, where $\langle \sigma v\rangle=a+b v^2+\mathcal{O}(v^4)$, and $a,b$ correspond, respectively, to the $s$- and $p$-wave components. Then, the constant DM annihilation to neutrinos essentially indicates $a\gg b$. When $|a|\sim |b|$ with $b<0$, however, there may exist certain cancellation between the $s$- and $p$-wave contributions, such that the annihilation cross section at freeze-out with $v=\mathcal{O}(0.1)$ could be smaller than that at the present day with $v=\mathcal{O}(10^{-4})$. Nevertheless, a difference at orders of magnitude for $\langle \sigma v\rangle$ between the freeze-out and present-day epochs, which is needed for observations from upcoming neutrino detectors,  requires a delicate fine-tuning between $a$ and $b$. One such tuning may arise in the resonant production regime. In this case, DM annihilation to neutrinos occurs through an $s$-channel  process, and the DM mass should be confined to nearly half of the mediator mass.

\section{Illustrative mechanisms for intermediate DM production}
\label{sec:check_inter}

The analysis presented in the previous sections relies on a phenomenological parameterization of the intermediate DM abundance, which allows us to investigate the resulting neutrino and CMB signatures in a model-independent manner. In this section, we discuss an illustrative example for the intermediate DM  production, based on the superweakly interacting massive particle (super-WIMP) mechanism~\cite{Feng:2003xh,Feng:2003uy}, and discuss how to deplete the additional DM  abundance, as well the potential challenges.

\subsection{Radiation domination}
\label{subsec:benchmark-radiation}

We first examine whether the energy density of the intermediate DM component remains subdominant during radiation domination.
For definiteness, let us consider a long-lived hidden scalar $\phi$, called super-WIMP, which is heavier than the DM particle $\chi$ and decays after neutrino decoupling via the two-body channel $\phi\to \chi+ \chi$. For simplicity, we take $\chi$ as a Majorana fermion. When the decay occurs at some temperature $T_{\rm dec}$, the energy density stored in $\phi$ is converted into the DM sector, giving
\begin{align}
Y_\chi=\frac{m_\phi Y_\phi}{E_\chi}\,.
\end{align}
The intermediate DM yield therefore directly reflects the abundance of the parent particle, $Y_\phi$.

To estimate whether the parent population causes an early matter-dominated epoch, we normalize its energy density to the background photon energy density:
\begin{align}
R_\rho\equiv\frac{\rho_\phi}{\rho_\gamma}
\approx 0.19 \, \left(\frac{m_\phi}{m_\chi}\right)
\left(\frac{f_\chi}{100}\right)
\left(\frac{g_s(T_{\rm dec})}{3.34}\right)
\left(\frac{1~\mathrm{keV}}{T_{\rm dec}}\right) \,,
\end{align}
where the parameterization of $Y_\chi$ in Eq.~\eqref{eq:YDM_param} has been used and $m_\phi/m_\chi>2$. Requiring $R_\rho<1$ gives
\begin{align}\label{eq:fchi_bound}
 f_\chi< 513.9
 \left(\frac{m_\chi}{m_\phi}\right)
 \left(\frac{T_{\rm dec}}{1~\mathrm{keV}}\right) \,.
\end{align}
For super-WIMP decay occurring at the keV temperatures, $T_{\rm dec}\sim 1~\mathrm{keV}$, the upper limit of Eq.~\eqref{eq:fchi_bound} can be approximately $f_\chi \sim 250$ when $m_\chi\to m_\phi/2$. Moreover,
because $R_\rho\propto T^{-1}$ before the decay, if $\rho_\phi<\rho_\gamma$ is satisfied at $T_{\rm dec}$, the  ratio $R_\rho$ is even smaller at earlier times. Thus, a benchmark value $f_\chi=100$ shown in Fig.~\ref{fig:sigmav-mX} can remain compatible with radiation domination for an appropriate parent-to-DM mass ratio.

\subsection{Depletion of the enhanced abundance}
\label{subsec:benchmark-depletion}

Since we assume that the dominant annihilation channel of DM is through SM neutrinos, 
we next ask whether the same channel will also help DM deplete the intermediate abundance to match the present-day value.
To see this, let us consider the Boltzmann evolution of DM from being created at $T_{\rm dec}$ with a fixed  $f_\chi$ down to the matter-radiation equality $T_{\rm eq}\simeq 2$~eV.  The Boltzmann equation for $Y_\chi=n_\chi/s$ can be written as
\begin{align}
\frac{d Y_\chi}{d z}=-\frac{\langle\sigma v\rangle s Y_\chi^2}{Hz}\,,
\label{eq:dYchi/dz}
\end{align}
where we defined the dimensionless temperature variable $z\equiv  T_{\rm eq}/T$ such that the initial condition gives \begin{align}
Y_\chi(z_i=T_{\rm eq}/T_{\rm dec})= 4.37\times 10^{-7}\left(\frac{1~\mathrm{MeV}}{m_{\chi}}\right) 
\left( f_\chi + 1 \right)\,.
\label{eq:dYchi-dz}
\end{align}
inherited from the super-WIMP decay, while the asymptotic condition matches the present-day value at $z = z_{\rm eq}=1$ with the corresponding $Y_\chi$ obtained by taking $f_\chi=0$ in Eq.~\eqref{eq:dYchi-dz}.  

The solution of Eq.~\eqref{eq:dYchi/dz} can be analytically obtained as
\begin{align}
Y_\chi(z)
&=c_{\rm eq}\frac{Y_{\chi,i}zz_i}
{c_{\rm eq} zz_i+Y_{\chi,i}\langle\overline{\sigma v}\rangle(z-z_i)}\,,
\end{align}
with $c_{\rm eq}=3.67\times10^{-3}$, $Y_{\chi,i} \equiv Y_\chi(z_i)$, $\langle\overline{\sigma v}\rangle \equiv \langle\sigma v\rangle/ (10^{-24}~\mathrm{cm}^3\,\mathrm{s}^{-1})$. At $z_{\rm eq}=1\gg z_i$, a small cross section leaves $Y_\chi$ close to its initial enhanced value $Y_{\chi}(z_{\rm eq})\approx Y_{\chi,i}$.
Efficient depletion through the neutrino channel requires the $\langle\sigma v\rangle$ term to dominate in the denominator.
Therefore, $Y_\chi(z_{\rm eq})\approx c_{\rm eq} z_i \langle\overline{\sigma v}\rangle^{-1}$, matching this value to the present-day one, i.e.,  Eq.~\eqref{eq:dYchi-dz}  with $f_\chi=0$, requires approximately
\begin{align}
\langle\sigma v\rangle
\simeq 8.41
\left(\frac{m_\chi}{1~\mathrm{MeV}}\right)
10^{-24}~\mathrm{cm}^3\,\mathrm{s}^{-1} \,.
\label{eq:rate_ann_eq}
\end{align}
For the masses of interest, this generally requires cross sections of order $10^{-22}~\mathrm{cm}^3\,\mathrm{s}^{-1}$ or larger, which are already excluded by current observations. The neutrino channel therefore cannot, by itself, provide the required late depletion in the relevant parameter space.


The large annihilation cross section required for this depletion also imposes a nontrivial condition on the intermediate DM abundance. If the corresponding hidden annihilation channel is already active when the enhanced population is produced, its rate relative to the Hubble expansion is
\begin{align}
\frac{\Gamma_{\rm ann}}{H}
\equiv
\frac{n_{\chi}\langle\sigma v\rangle}{H}
\simeq
1.86\,(1+f_{\chi})
\left(\frac{T}{1~\mathrm{keV}}\right),
\end{align}
where Eqs.~\eqref{eq:dYchi-dz} and \eqref{eq:rate_ann_eq} have been used.
For a large intermediate abundance, such as the benchmark $f_{\chi}=100$ adopted in Fig.~\ref{fig:sigmav-mX}, 
one has $\Gamma_{\rm ann}\gg H$ at keV temperatures, so that annihilation can efficiently deplete the produced DM. 
Sustaining such an abundance therefore requires certain balance between the production through $\phi\to\chi\chi$ and the depletion.
Achieving this balance over an extended period depends on the detailed production and annihilation dynamics, which can make the assumed abundance plateau challenging to realize in a concrete model.


If DM annihilates to some hidden particles with a large annihilation cross section, most of the DM abundance created in the intermediate stage will be converted to the hidden particle. 
Since the annihilation cross section is large, we may estimate the energy stored in the hidden particle by following the super-WIMP mechanism. 
In this case, most of the intermediate DM abundance is converted into a hidden particle $\eta$. The energy stored in this population is estimated by
\begin{align}
\rho_\eta=E_{\eta,i}Y_{\eta, i} s_i
\approx \rho_\chi(\text{intermediate})
\approx m_{\chi} Y_{\chi,i} s_i  \,.
\end{align}
The initial characteristic energy of $\eta$ is taken as $E_{\eta,i} \simeq m_\chi$. After production, let us suppose that $\eta$ is decoupled and remains relativistic. Its energy then simply redshifts as $E_\eta=m_\chi T/T_{\rm dec}$.
If $\eta$ remains relativistic until matter-radiation equality $z_{\rm eq}=1$, we can estimate its contribution to $N_{\rm eff}$ that will be measured by CMB experiments, with
\begin{align}
N_{\rm eff}^{\eta}
=\frac{8}{7}\left(\frac{11}{4}\right)^{4/3}
\frac{Y_{\chi,i}m_\chi T_{\rm CMB} s_{\rm CMB}}{T_{\rm dec}\rho_{\gamma,{\rm CMB}}}
\approx0.43
\left(\frac{f_\chi}{100}\right)
\left(\frac{1~\mathrm{keV}}{T_{\rm dec}}\right) \,.
\end{align}
Thus, for $f_\chi\simeq100$ and $T_{\rm dec}$ near the keV scale, genuinely dark-radiation final states will produce an additional $N_{\rm eff}$ contribution, significantly modifying the cosmological predictions shown in Fig.~\ref{fig:sigmav-mX}.

This conclusion can be avoided, for instance, if $\eta$ is heavier such that it is relativistic immediately after the DM annihilation but becomes nonrelativistic before the matter-radiation epoch. In this case, $\eta$ does not contribute to $N_{\rm eff}$  at the CMB epoch.  This requires the momentum $p_\eta=m_\chi T_{\rm eq}/T_{\rm dec}<m_\eta$, or 
\begin{align}
m_\eta>m_\chi\left(\frac{10^{-3}~\mathrm{keV}}{T_{\rm dec}}\right) \,.
\label{eq:mass-relation}
\end{align}
At $T_{\rm eq}$, the rest-mass energy density of $\eta$, normalized to the photon energy density, is approximately
\begin{align}
R_\eta\equiv\frac{\rho_\eta}{\rho_\gamma}
\approx0.1
\left(\frac{f_\chi}{100}\right)
\left(\frac{10^3m_\eta}{m_\chi}\right)
> 0.1
\left(\frac{f_\chi}{100}\right)
\left(\frac{1~\mathrm{keV}}{T_{\rm dec}}\right) \,.
\end{align}
Since the DM and radiation energy densities should be  comparable around the matter-radiation equality epoch,
$\rho_\gamma\sim\rho_\chi$, $R_\eta$ provides an order-of-magnitude estimate for the fractional matter-density
contribution from a nonrelativistic $\eta$ population. 
Requiring the relic $\eta$ population to remain a subdominant component of the DM density, which we estimate by imposing
$R_\eta\lesssim0.1$, gives the approximate upper bound
\begin{align}
f_\chi \lesssim 100\left(\frac{T_{\rm dec}}{1~\mathrm{keV}}\right) \,.
\end{align}

In summary, an intermediate DM abundance may be depleted through annihilation into some hidden particles, which are relativistic immediately after DM annihilation but subsequently become nonrelativistic before the matter-radiation epoch owing to the redshift effect.

\subsection{Absence of chemical recoupling}
\label{subsec:benchmark-recoupling}

The intermediate DM production under consideration occurs at a low temperature where $T_{\rm dec}\ll m_\chi$.
Our analysis assumes that the subsequent DM annihilation can be treated as an effectively one-way process,
$\chi+\bar\chi\to\nu+\bar\nu$, with the produced neutrinos subsequently free-streaming. If the inverse reaction became efficient to chemically recouple the injected neutrinos with DM, this approximation would no longer be valid: the backward process would partially replenish the DM population, modify its depletion history, and alter the resulting nonthermal neutrino flux. It is therefore necessary to verify that such chemical recoupling does not occur in the parameter region considered.

To identify the potentially relevant backward processes, it is useful to distinguish the inverse-annihilation channels according to whether the participating neutrinos belong to the thermal background or to the nonthermal population injected by DM annihilation. It is easy to understand that 
the backward processes $\nu_{\rm bg}+\bar\nu_{\rm bg}\to\chi+\bar\chi$ and $\nu_{\rm bg}+\bar\nu_{\rm inj}\to\chi+\bar\chi$ are strongly suppressed by the kinematic threshold, since the background neutrino $\nu_{\rm bg}$ has an energy at the temperature scale much lower than the DM mass.  
These two inverse processes are therefore kinematically forbidden for
neutrinos with their characteristic energies. The only inverse-annihilation channel that could potentially compete
with the forward reaction is therefore
$\nu_{\rm inj}+\bar\nu_{\rm inj}\to\chi+\bar\chi$.
The relevant criterion is that the backward collision term remains negligible,
\begin{align}
{\cal B}(T)
\equiv
\frac{{\cal C}_{\nu_{\rm inj}\bar\nu_{\rm inj}\to\chi\bar\chi}}
     {{\cal C}_{\chi\bar\chi\to\nu_{\rm inj}\bar\nu_{\rm inj}}}
=
\kappa_{\rm eff}(T)
\left(\frac{n_\nu^{\rm inj}}{n_\chi}\right)^2
\ll 1  \,,
\end{align}
where $\kappa_{\rm eff}(T)
\equiv\langle\sigma v\rangle_{\nu_{\rm inj}\bar\nu_{\rm inj}\to\chi\bar\chi}
/
\langle\sigma v\rangle_{\chi\bar\chi\to\nu_{\rm inj}\bar\nu_{\rm inj}}$ 
is evaluated by averaging over the actual nonthermal distribution of
the injected neutrinos.
In the nonrelativistic approximation adopted here, each neutrino is injected with a characteristic energy $E_\nu^{\rm inj}\simeq m_\chi$.
However, as shown in Eq.~\eqref{eq:f_nu_sol_1}, the injected-neutrino distribution decreases toward the high-energy end,
$f_\nu^{\rm inj}\propto (E_\nu^{\rm inj})^{-4}$.
Consequently,  phase-space configurations capable of approaching the kinematic threshold for the inverse reaction are strongly suppressed.
This leads to a strongly suppressed inverse rate, implying $\kappa_{\rm eff}(T)\ll1$. For completeness, it is also useful to estimate the accumulated injected-neutrino abundance relative to the DM abundance. 
For one neutrino flavor, evaluated at the end of the intermediate production stage, we find
\begin{align}
\frac{n_\nu^{\rm inj}}{n_\chi}
\simeq
0.11
\left(
\frac{\langle\sigma v\rangle}
     {10^{-24}~{\rm cm}^3\,{\rm s}^{-1}}
\right)
\left(\frac{1~{\rm MeV}}{m_\chi}\right)
(f_\chi+1)\,,
\end{align}
which, for masses and cross section of interest, is indeed small even for $f_\chi\simeq 100$, and hence ${\cal B}(T)\ll 1$ is justified.

As a complementary check, one can also compare the forward neutrino-production source with the Hubble-dilution term in the Boltzmann equation for the neutrino number density. We define the ratio
\begin{align}
R_{\rm th}
\equiv
\frac{
\langle\sigma v\rangle n_{\chi,i}^{2}
}{
3Hn_\nu^{\rm eq}
} \,,
\end{align}
where
\begin{align}
n_\nu^{\rm eq} = \frac{3\zeta(3)}{4\pi^2}T^3\,,
\end{align}
is the equilibrium number density of neutrinos. Using the DM abundance during the intermediate production stage, we obtain
\begin{align}
R_{\rm th}
\simeq
1.29\times10^{-5}
\left(\frac{f_\chi}{100}\right)^2
\left(\frac{20~{\rm MeV}}{m_\chi}\right)^2
\left(\frac{T}{1~{\rm keV}}\right)
\left(
\frac{\langle\sigma v\rangle}
     {10^{-24}~{\rm cm}^3\,{\rm s}^{-1}}
\right) \,.
\end{align}
The quantity $R_{\rm th}$ measures whether the forward annihilation
 can generate an equilibrium-sized neutrino abundance within
approximately one Hubble time.
For the low-mass benchmark region, $m_\chi\sim20$~MeV and $\langle\sigma v\rangle\lesssim10^{-25}~\mathrm{cm}^3\,\mathrm{s}^{-1}$, one finds that  $R_{\rm th}\ll1$ for $T\sim 1$~keV and $f_\chi=100$.  The same remains true for heavier DM around $m_\chi\sim1$~GeV even when the relevant cross section reaches $\sim10^{-23}~\mathrm{cm}^3\,\mathrm{s}^{-1}$.  In fact, requiring $R_{\rm th}<1$ gives, for representative low-mass parameters,
\begin{align}
f_\chi\lesssim10^5 \,,
\end{align}
which is far weaker than the constraint obtained from Eq.~\eqref{eq:fchi_bound}.
Therefore, DM re-thermalization with neutrinos does not occur under the benchmark scenario shown in Fig.~\ref{fig:sigmav-mX}.

\subsection{Dependence on the peak abundance}
\label{subsec:fchi-dependence_50}

The benchmark scenario shown in Fig.~\ref{fig:sigmav-mX} is obtained by taking $f_\chi=100$. From previous discussions based on the super-WIMP mechanism, we illustrated that $f_\chi=100$ is a reasonable value for the intermediate DM production. Nevertheless, it is also interesting to see 
 how the joint probe from cosmology and neutrino detectors depends on the intermediate DM abundance. To this end,  we repeat the calculation by taking a more conservative value $f_\chi=50$, while keeping $T_{\rm pro}=0.5~\mathrm{keV}$ and $T_{\rm end}=0.1~\mathrm{keV}$ fixed. 

\begin{figure}[t]
\centering
\includegraphics[scale=0.41]{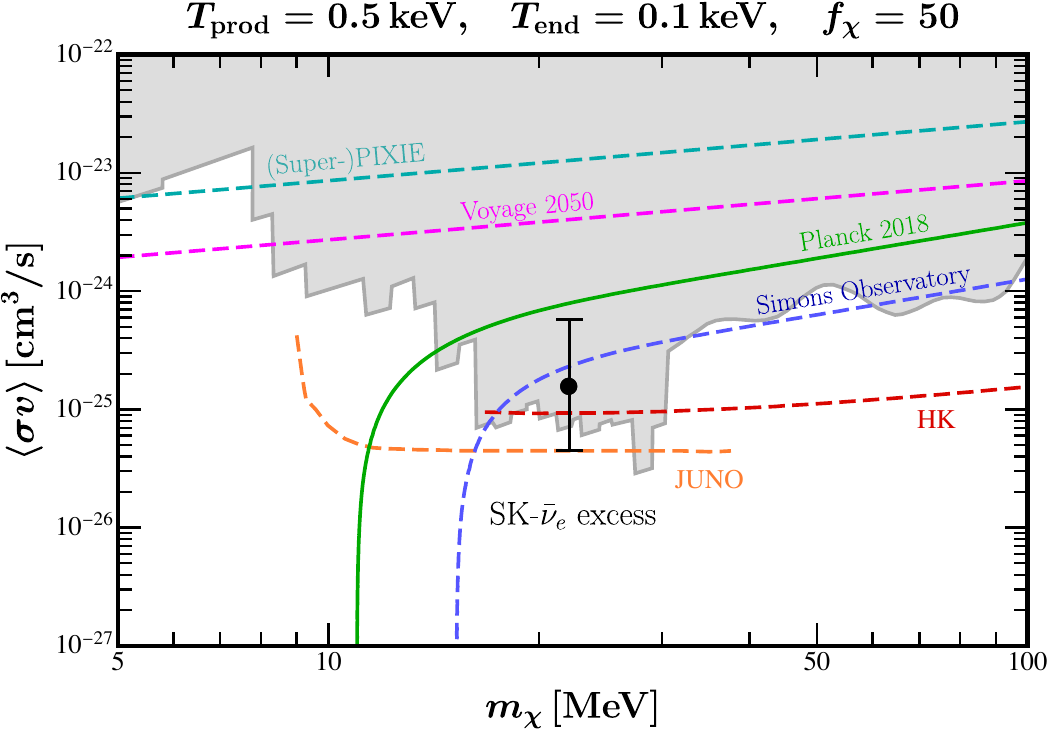}
\quad
\includegraphics[scale=0.41]{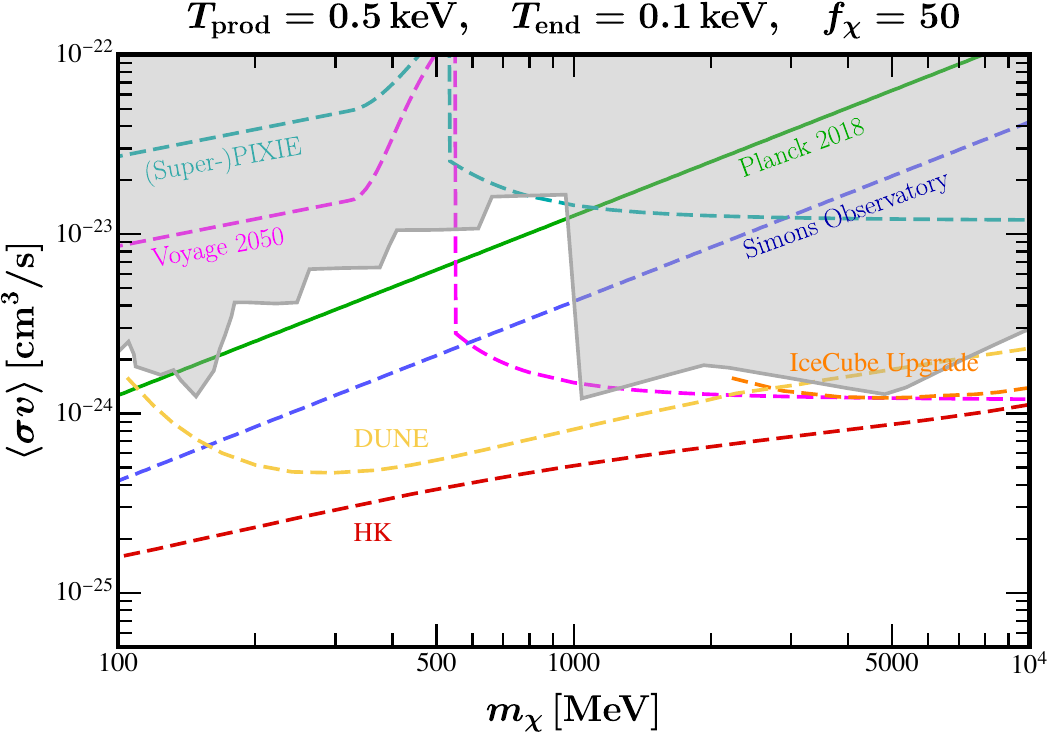}
\caption{Detection windows in the
$(m_\chi,\langle\sigma v\rangle)$ plane for the reduced peak intermediate abundance $f_\chi=50$, with $T_{\rm pro}=0.5~\mathrm{keV}$ and $T_{\rm end}=0.1~\mathrm{keV}$.
The left and right panels show the low- and high-mass regions, respectively. 
The cosmological target contours are recalculated for the reduced intermediate abundance, whereas the neutrino detector sensitivities, current limits, and the SK-$\bar\nu_e$ excess region are unchanged from Fig.~\ref{fig:sigmav-mX}.}
\label{fig:sigmav-fchi50}
\end{figure}

The displacement of the cosmological contours follows from the abundance dependence of the corresponding signals, which can be seen from the following  scaling 
\begin{align}
N_{\rm eff}^{\rm inj} \propto f_\chi(f_\chi+2)\langle\sigma v\rangle \,,
\quad
\mu_{\rm co} \propto f_\chi^2 \,,
\quad
\mu_{\rm pair} \propto f_\chi^4\,,
\end{align}
at a fixed $(m_\chi,\langle\sigma v\rangle)$ point. In particular, 
reducing $f_\chi$ from $100$ to $50$ suppresses the contributions of $N_{\rm eff}^{\rm inj}$ and $\mu_{\rm co}$ by a factor of four while the suppression for $\mu_{\rm pair}$ is even stronger.

In the low-mass panel of Fig.~\ref{fig:sigmav-fchi50}, the Planck 2018 and Simons Observatory contours move to larger annihilation cross sections relative to the $f_\chi=100$ result, while preserving their overall shapes. 
The same upward displacement occurs for the CMB $\mu$ distortion contours. 
Around the SK-$\bar\nu_e$ best-fit mass, the Simons Observatory contour lies above the central best-fit
point for $f_\chi=50$, though part of the quoted uncertainty region still extends into the cosmological target range. 
The CMB $\mu$ distortion contours remain well separated from the MeV-scale neutrino detector sensitivities, so the conclusion that a detectable $\mu$ distortion is not expected in the low-mass region is unchanged.

In the high-mass panel, lowering $f_\chi$ likewise shifts the cosmological target contours toward larger annihilation cross sections.
The pronounced change in the CMB $\mu$ distortion contours near $m_\chi\simeq500~\mathrm{MeV}$ remains at approximately the same mass: its location is controlled primarily by the kinematic onset of coannihilation with the relic neutrino background, rather than by the normalization of the intermediate abundance. 
Above this threshold, the coannihilation contribution becomes increasingly important, and the $\mu$ distortion contours depend only weakly on $m_\chi$. The benchmark with $f_\chi=50$ therefore preserves the characteristic high-mass window for cosmological probe, but moves its overlap with DUNE, HK, and IceCube Upgrade to larger values of $\langle\sigma v\rangle$.

In summary, here we adopt a more conservative benchmark $f_\chi=50$ to illustrate how the parameter space accessible to the joint probe from cosmology and neutrino detectors varies relative to $f_\chi=100$ shown in Fig.~\ref{fig:sigmav-mX}. Since $f_\chi$ encodes the underlying physics responsible for the intermediate DM production, different values of $f_\chi$ may lead to distinctive signals in various experiments. The joint-probe strategy proposed in this work may therefore shed some light on the underlying physical origin of the intermediate DM production required to resolve the density-deficit problem in realistic DM models.

\section{Conclusion}
\label{sec:con}
If DM dominantly annihilates into the SM neutrinos, it can partly explain the current null results of indirect DM detection and evade the more severe constraints from electromagnetic or hadronic energy injection in astrophysics and cosmology. Nevertheless, the weak interactions of SM neutrinos also present challenges for indirectly detecting this less constrained channel. Strikingly, several neutrino detectors, such as SK, JUNO, HK, and DUNE, will be able to detect extra neutrino fluxes potentially from DM annihilation, especially given the current hint of the SK antineutrino excess. To detect and confirm the neutrino signals, however, it would introduce the DM density-deficit problem: there must be some intermediate DM production processes to compensate for the over-annihilated DM density at freeze-out.  

In this work, we have investigated the consequences of resolving the density-deficit problem by combining the detection sensitivities from terrestrial neutrino detectors with the cosmic probe from CMB experiments. Instead of building the specific particle-physics models where DM can dominantly annihilate into neutrinos, we adopt a simple strategy, which is model independent and a rather sensible parameterization, to simulate the cosmic impacts from large rates of MeV-GeV DM annihilation into neutrinos. We demonstrate that there exists a large detection overlap between neutrino detectors and CMB experiments targeting the early Universe, including the probes of the CMB $\mu$ distortion and $N_{\rm eff}$. The results presented in this work are conservative in the sense that the cosmic impacts are entirely from DM annihilation to neutrinos, the same quantity that is probed by neutrino detectors. In specific particle-physics models, on the other hand, we do not exclude any possibility that may contribute additionally to the cosmic observables under consideration. In this situation, looking for DM annihilation to neutrinos complementarily through neutrino detectors and cosmic probes proposed in this work will be strengthened.

\section*{Acknowledgements}
We would like to thank John Beacom for discussions at the early stage of this project. We also thank Yushi Mura for discussions on the potential UV framework for the DM density-deficit problem, and Alessandro Granelli, Silvia Pascoli, and Salvador Rosauro-Alcaraz for explanations of the SK antineutrino excess.

\appendix

\section{Analytic derivation of injected neutrino distribution  and extra radiation}
\label{app:analy_sol}

In this appendix, we provide the explicit derivation of the analytic results used in Sections~\ref{sec:neu_inj} and \ref{sec:N_eff}. We begin by deriving the distribution function $f_{\nu}^{\rm inj}$ for neutrinos injected by DM annihilation after neutrino decoupling. We then compute the radiation energy density induced by this nonthermal component and use it to obtain the associated contribution to the effective number of neutrino species, $N_{\rm eff}^{\rm inj}$.

\subsection{Injected neutrino distribution $f_{\nu}^{\rm inj}$}

We first derive the collision term $\mathcal{C}_{\nu}$ that appears in the Boltzmann equation~\eqref{eq:nu_Boltzmann}. For neutrino production from DM annihilation, $\chi\bar{\chi}\to\nu\bar{\nu}$, the collision term is constructed directly as
\begin{align}
\mathcal{C}_{\nu}
=\frac{1}{2E_{\nu}}\int \left(\prod_{i=\chi, \bar{\chi}, \bar{\nu}} d\Pi_{i} \right) (2\pi)^{4} \delta^{4}(p_{\chi}+p_{\bar{\chi}}-p_{\nu}-p_{\bar{\nu}}) |\mathcal{M}|^{2} f_{\chi} f_{\bar{\chi}}
\, .
\end{align}
After performing the integration over the momentum space,  we obtain
\begin{align}
\mathcal{C}_{\nu}
&=\frac{1}{2E_{\nu}}\left(\int \frac{d^3 p_{\chi}}{(2\pi)^{3} } f_{\chi} \right)\left(\int \frac{d^3 p_{\bar{\chi}}}{(2\pi)^{3} }f_{\bar{\chi}}\right) \frac{\pi}{4E_{\bar{\nu}}E_{\chi}E_{\bar\chi}} \delta(E_{\chi}+E_{\bar{\chi}}-E_\nu-E_{\bar{\nu}}) |\mathcal{M}|^{2}
\nonumber \\[0.2cm]
&\approx \frac{\pi |\mathcal{M}|^{2}}{16 E_{\nu}^{2} m_{\chi}^{2}}\delta(E_{\nu}-m_{\chi}) n_{\chi}^{2}\label{eq:Cnu-2}
\, ,
\end{align}
where we take the approximation of nonrelativistic DM such that the squared amplitude $|\mathcal{M}|^2$ is both temperature and momentum independent, and the DM distribution functions can be integrated out to the number density, giving $n_{\chi}^{2}$ for symmetric DM $n_\chi=n_{\bar\chi}$.  
In the nonrelativistic limit, $E_\chi=E_{\bar{\chi}}\simeq m_\chi$ is used, while the energy conservation condition $E_\nu=E_{\bar \nu}\approx m_\chi$  is  time dependent due to cosmic expansion.

For the $s$-wave nonrelativistic DM annihilation, there is a simple relation between the squared amplitude $|\mathcal{M}|^{2}$ and the thermally averaged cross section $\langle \sigma v\rangle$, the latter of which is defined through the Boltzmann collision rate of the injected neutrino number density
\begin{align}
 \langle \sigma v\rangle n_\chi^2\equiv  \int \left(\prod_{i=\chi, \bar{\chi},\nu, \bar{\nu}} d\Pi_{i} \right) (2\pi)^{4} \delta^{4}(p_{\chi}+p_{\bar{\chi}}-p_{\nu}-p_{\bar{\nu}}) |\mathcal{M}|^{2} f_{\chi} f_{\bar{\chi}}\,.
\end{align}
Using a constant $|\mathcal{M}|^2$, the above integration over the phase space can be performed independently of $|\mathcal{M}|^2$, such that 
\begin{align}
\langle \sigma v\rangle \approx \frac{|\mathcal{M}|^{2}}{32\pi m_{\chi}^{2}} \,.
\end{align}
The collision term $\mathcal{C}_{\nu}$ given in Eq.~\eqref{eq:Cnu-2} can then be expressed in terms of $\langle \sigma v\rangle$
\begin{align}
\mathcal{C}_{\nu}=\frac{2\pi^{2}\langle \sigma v\rangle }{E_{\nu}^{2}}n_{\chi}^{2} \delta(E_{\nu}-m_{\chi})\,.
\end{align}

We now solve the Boltzmann equation for the injected neutrino distribution function
\begin{align}
\frac{d f_{\nu}^{\rm inj}}{d  t} =\frac{\partial f_{\nu}^{\rm inj}}{\partial t}-H|\p|\frac{\partial f_{\nu}^{\rm inj}}{\partial |\p|}=\mathcal{C}_{\nu} \,.
\end{align}
The Liouville total derivative is taken along the phase-space trajectory of a freely propagating neutrino. In an expanding universe, the physical momentum redshifts as
\begin{align}
\frac{d |\p|}{d t} = -H |\p| \,,
\end{align} 
or equivalently, the comoving momentum $|\q|\equiv a|\p|$ is conserved. Therefore, the dispersion relation of the injected neutrino reads
\begin{align}
E_{\nu}(t') = |\p| \frac{a(t)}{a(t')} \, ,
\quad
\dot{E}_{\nu}(t') = -|\p| \frac{a(t) \dot{a}(t')}{a^{2}(t')} = -E_{\nu}(t') H(t') \, .
\end{align}
Imposing the initial condition $f_{\nu}^{\rm inj}(t_{\rm pro},|\p|)=0$, the solution is obtained by integrating the collision term along the phase-space characteristic:
\begin{align}
f_{\nu}^{\rm inj}(t, |\p|) &= \int_{t_{\rm pro}}^{t} dt' \, \mathcal{C}_{\nu} (t',|\p|(t'))  
\nonumber \\[0.2cm]
&= \int_{t_{\rm pro}}^{t} dt' \, \frac{2\pi^{2}\langle \sigma v\rangle }{E_{\nu}^{2}(t')}n_{\chi}^{2}(t') \delta(E_{\nu}(t')-m_{\chi})
\nonumber \\[0.2cm]
&= \frac{2\pi^{2}\langle \sigma v\rangle }{E_{\nu}^{2}(t_{*})}n_{\chi}^{2}(t_{*}) \left|\dot{E}_{\nu}(t_{*})\right|^{-1}
\theta(t_{*}-t_{\rm pro})\theta(t -t_{*})
\nonumber \\[0.2cm]
&= \frac{2\pi^{2}\langle \sigma v\rangle }{m_{\chi}^3  H(t_{*})}n_{\chi}^{2}(t_{*}) \theta(t_{*}-t_{\rm pro})\theta(t -t_{*})
\, ,
\end{align}
where $t_{*}$ is defined by
\begin{align}
E_{\nu}(t_{*}) = m_{\chi} \,.
\end{align}
The factor $|\dot{E}_\nu(t_{*})|^{-1}$ is the Jacobian obtained when the energy  $\delta$-function is converted into a time-dependent one. The two Heaviside step functions enforce that $t_{*}$ lies inside the integration range $t_{\rm pro}<t_{*}<t$.

To make the redshift dependence explicit, it is convenient to trade the dimensional variables for dimensionless combinations. Since both the neutrino energy and the cosmic temperature scale approximately as $a^{-1}$ during the epoch of interest, we therefore introduce the dimensionless variables
\begin{align}
r \equiv \frac{E_{\nu}}{T} \, ,
\quad
x \equiv \frac{T_{\rm pro}}{T} \, ,
\quad
r_{\chi} \equiv \frac{m_{\chi}}{T_{\rm pro}} \, .
\end{align}
The resonant condition $E_\nu(t_{*})=m_\chi$ then fixes the corresponding temperature as
\begin{align}
T_{*}=\frac{m_\chi}{r}\, .
\end{align}
The two Heaviside functions appearing in the solution can be rewritten in terms of these dimensionless variables as
\begin{align}
\theta(r-r_{\chi}) 
&= \theta \left(\frac{E_{\nu}}{T}-\frac{m_{\chi}}{T_{\rm pro}} \right)
= \theta \left(E_{\nu}-\frac{T}{T_{\rm pro}}m_{\chi} \right)
\nonumber \\[0.2cm]
&= \theta \left(E_{\nu}(t_{\rm pro})-m_{\chi} \right)
= \theta \left(E_{\nu}(t_{\rm pro})-E_{\nu}(t_{*}) \right)
= \theta \left(t_{*} - t_{\rm pro} \right)
\, ,
\\[0.2cm]
\theta(r_{\chi}x -r) 
&= \theta \left(\frac{m_{\chi}}{T}-\frac{E_{\nu}}{T} \right)
= \theta \left(m_{\chi} - E_{\nu} (t) \right)
\nonumber \\[0.2cm]
&= \theta \left(E_{\nu}(t_{*}) - E_{\nu} (t) \right)
= \theta \left(t - t_{*} \right)
\, .
\end{align}
These relations show that the condition $t_{\rm pro}<t_{*}<t$ is equivalent to the finite support
\begin{align}
r_{\chi}\leq r\leq r_{\chi}x\, .
\end{align}
Using
\begin{align}
H(T_{*})=1.66\sqrt{g_{\rho}(T_{*})}\frac{T_{*}^{2}}{M_{\rm P}}\, ,
\quad
n_\chi(T_{*})=s(T_{*})Y_\chi(T_{*})\, ,
\quad
s(T_{*})=\frac{2\pi^{2}}{45} g_{s}(T_{*}) T_{*}^{3}\, ,
\end{align}
the injected neutrino distribution function becomes
\begin{align}
f_{\nu}^{\rm inj}(t, |\p|) 
&= \frac{2\pi^{2}\langle \sigma v\rangle }{m_{\chi}^3  H(T_{*})} s^{2}(T_{*}) Y^{2}_{\chi}(T_{*}) \theta(r-r_{\chi})\theta(r_{\chi}x -r)
\nonumber \\[0.2cm]
&= \left(\frac{2\pi^{2}}{45}\right)^{2} \frac{2\pi^{2} M_{\rm P} \langle \sigma v\rangle }{m_{\chi}^3  1.66\sqrt{g_{\rho}(T_{*})}}  g^{2}_{s}(T_{*}) \left( \frac{m_{\chi}}{r}\right)^{4} \theta(r-r_{\chi})\theta(r_{\chi}x -r)
\nonumber \\[0.2cm]
&\quad \times \left(4.37\times 10^{-7}\left(\frac{1~\mathrm{MeV}}{m_{\chi}}\right)\left[f_{\chi} \theta(T_{\rm pro}-T_{*})\theta(T_{*}-T_{\rm end})+1\right]\right)^{2}
\nonumber \\[0.2cm]
&= \left(\frac{2\pi^{2}}{45}\right)^{2} 
\frac{g^{2}_{s}(T_{*})}{1.66\sqrt{g_{\rho}(T_{*})}}
\frac{2\pi^{2} M_{\rm P} \langle \sigma v\rangle T_{\rm pro}^5 r_{\chi}^{3}}{m_{\chi}^{4} r^{4}} \theta(r-r_{\chi})\theta(r_{\chi}x -r)
\nonumber \\[0.2cm]
&\quad \times \left(4.37\times 10^{-7}\left(\frac{1~\mathrm{MeV}}{10~\mathrm{keV}} \frac{10~\mathrm{keV}}{T_{\rm pro}}\right)\left[f_{\chi} \theta(r-r_{\chi})\theta(T_{\rm pro}r_{\chi}-T_{\rm end}r)+1\right]\right)^{2}
\nonumber \\[0.2cm]
&= \frac{2\pi^{2} M_{\rm P}^{*} \langle \sigma v\rangle T_{\rm pro}^{2}}{m_{\chi} r^{4}} \Theta^{2} \theta(r-r_{\chi})\theta(r_{\chi}x -r)
\, ,
\end{align}
where we have defined notations
\begin{align}
M_{\rm P}^{*} &\equiv 0.437^{2} \left(\frac{2\pi^{2}}{45}\right)^{2} \frac{g^{2}_{s}(T_{*})}{1.66\sqrt{g_{\rho}(T_{*})}} M_{\rm P} \approx \frac{0.45}{g_{\rho}(T_{*})} M_{\rm P} \approx 0.135 \, M_{\rm P}  \, ,
\\[0.2cm]
\Theta &\equiv 10^{-4}
\left(\frac{10~{\rm keV}}{T_{\rm pro}}\right)
\left[f_{\chi}\theta(r-r_{\chi})\theta(T_{\rm pro}r_{\chi}-T_{\rm end}r)+1\right] \, .
\end{align}
After neutrino decoupling, we can take $g_{s}(T)\approx g_{\rho}(T)\approx 3.34$, which is a good approximation over the temperature range considered here. The factor $\Theta$ encodes the time dependence of the DM abundance: the term proportional to $f_{\chi}$ selects annihilation during the intermediate production stage, while the constant term corresponds to annihilation from the final relic abundance.

\subsection{Energy density and contribution to $N_{\rm eff}$}

We next compute the energy density carried by the injected neutrinos. This step connects the nonthermal neutrino spectrum derived above to the cosmological observable $N_{\rm eff}$. Since these neutrinos are produced after neutrino decoupling and remain free streaming, their energy density behaves as an additional radiation component.

Assuming comparable rates of DM annihilation into the three SM neutrino flavors, the energy density stored in the injected neutrinos is
\begin{align}
\rho_{\nu}^{\rm inj}
&=3\int \frac{d^3 p}{(2\pi)^3} E_{\nu} f_{\nu}^{\rm inj}
 \\[0.2cm]
&=3\int \frac{|\p|^{2} d|\p| d\Omega}{(2\pi)^3} E_{\nu} f_{\nu}^{\rm inj}
\nonumber \\[0.2cm]
&=3 \, M_{\rm P}^{*} \langle \sigma v\rangle m_{\chi} T^{4} \times 10^{-12} \left(\frac{1~\mathrm{MeV}}{m_{\chi}}\right)^{2} 
\nonumber \\[0.2cm]
&\quad \times
\int dE_{\nu} \, \frac{1}{E_{\nu}} 
\biggl[f_{\chi}(f_{\chi}+2) \theta \left(\frac{T}{T_{\rm end}} m_{\chi} - E_{\nu}\right) + 1 \biggr] 
\theta \left(E_{\nu} - \frac{T}{T_{\rm pro}} m_{\chi} \right) \theta \left(m_{\chi} - E_{\nu} \right)\nonumber
\,.
\end{align}
The remaining integral can be calculated straightforwardly to yield $\mathcal{F}_{\rm int} + \mathcal{F}_{\rm rel}$, where  we have separated the result into the intermediate-production contribution and the relic contribution:
\begin{align}
\mathcal{F}_{\rm int}
&\equiv \theta \left(T_{\rm pro} - T \right) 
f_{\chi}(f_{\chi}+2) \biggl[ \theta \left(T - T_{\rm end} \right)  \ln \left( \frac{T_{\rm pro}}{T} \right)
+ \theta \left(T_{\rm end} - T \right) \ln \left( \frac{T_{\rm pro}}{T_{\rm end}} \right)
\biggr] 
\, ,
\\
\mathcal{F}_{\rm rel}
&\equiv \theta \left(T_{\rm pro} - T \right) \ln \left( \frac{T_{\rm pro}}{T} \right)
\, .
\end{align}
The injected neutrino energy density is therefore
\begin{align}
\rho_{\nu}^{\rm inj}
&=3 \, M_{\rm P}^{*} \langle \sigma v\rangle m_{\chi} T^{4} \times 10^{-12} \left(\frac{1~\mathrm{MeV}}{m_{\chi}}\right)^{2} 
\left(\mathcal{F}_{\rm int} + \mathcal{F}_{\rm rel} \right).
\end{align}
We next convert this energy density into the corresponding contribution $N_{\rm eff}^{\rm inj}$ with the photon energy density 
\begin{align}
\rho_{\gamma} = \frac{\pi^{2}}{15} T^{4} \, . 
\end{align}
Since $\rho_{\nu}^{\rm inj}$ denotes the injected neutrino contribution summed over three flavors, the corresponding antineutrinos give an additional factor of $2$. Thus,
\begin{align}
N_{\rm eff}^{\rm inj}
&=2 \times \frac{8}{7}\left(\frac{11}{4}\right)^{4/3}\left(\frac{\rho_{\nu}^{\rm inj}}{\rho_{\gamma}}\right)
\nonumber \\
&= 6 \times \frac{8}{7}\left(\frac{11}{4}\right)^{4/3} \frac{1.5 \times 10^{-11}}{\pi^{2}} M_{\rm P}^{*} \langle \sigma v\rangle \left(1~\mathrm{MeV} \right) \left(\frac{1~\mathrm{MeV}}{m_{\chi}}\right) 
\left(\mathcal{F}_{\rm int} + \mathcal{F}_{\rm rel} \right)
\, . 
\end{align}
Using $M_{\rm P}^{*}\simeq 0.135 \, M_{\rm P}$, one finds
\begin{align}
N_{\rm eff}^{\rm inj}
&\simeq 0.0056
\left(\frac{\langle\sigma v\rangle}{10^{-24}~\mathrm{cm}^3/\mathrm{s}}\right)
\left(\frac{1~\mathrm{MeV}}{m_{\chi}}\right)
\left(\mathcal{F}_{\rm int}+\mathcal{F}_{\rm rel}\right)
\, .
\end{align}
At the CMB epoch, $T=T_{\rm CMB}$, and for $T_{\rm CMB}<T_{\rm end}<T_{\rm pro}$, this becomes
\begin{align}
N_{\rm eff}^{\rm inj}
&\simeq 0.0056
\left(\frac{\langle\sigma v\rangle}{10^{-24}~\mathrm{cm}^3/\mathrm{s}}\right)
\left(\frac{1~\mathrm{MeV}}{m_{\chi}}\right)
\left[
f_{\chi}(f_{\chi}+2)\ln \left( \frac{T_{\rm pro}}{T_{\rm end}} \right)
+\ln \left( \frac{T_{\rm pro}}{T_{\rm CMB}} \right)
\right]
\, .
\end{align}
The first logarithm measures the duration of the intermediate stage, while the second logarithm is the relic contribution accumulated from $T_{\rm pro}$ to the CMB epoch. This logarithmic dependence explains why moderate variations of $T_{\rm pro}$ and $T_{\rm end}$ only mildly affect $N_{\rm eff}^{\rm inj}$.
The full result shows how the intermediate DM production required to resolve the density-deficit problem can enhance $N_{\rm eff}^{\rm inj}$ and thereby provide a complementary cosmological probe of the same annihilation process.
In the absence of intermediate DM production,  we take $\mathcal{F}_{\rm int}=0$, such that the expression reduces to
\begin{align}
N_{\rm eff}^{\rm inj}
&= 0.0056
\left(\frac{\langle\sigma v\rangle}{10^{-24}~\mathrm{cm}^3/\mathrm{s}}\right)
\left(\frac{1~\mathrm{MeV}}{m_{\chi}}\right)
\ln \left( \frac{T_{\rm pro}}{T_{\rm CMB}} \right)
\, ,
\end{align}
recovering the irreducible neutrino injection from the final relic abundance~\cite{Kanemura:2025byi}.

\bibliographystyle{JHEP}
\bibliography{Refs}

\end{document}